%% file: 160825_arxiv.tex
\documentclass[11pt,letter]{article}
\usepackage{graphicx}
\usepackage{setspace}
\pdfoutput=1
\usepackage{lmodern}
\usepackage{amsmath}
\usepackage{amsthm}{\theoremstyle{plain}\newtheorem{assumption}{Assumption}}
\usepackage[margin=3cm]{geometry}
\usepackage[round]{natbib}
\usepackage{bibentry}
\usepackage{appendix}
\usepackage{lmodern}
\usepackage{hyperref}
\usepackage{caption}
\newenvironment{appendixnumbering}{}{}

\title{THE INFORMATIONAL CONTENT OF THE LIMIT ORDER BOOK: AN EMPIRICAL STUDY OF PREDICTION MARKETS\thanks{Carnegie Mellon University, 5000 Forbes Avenue, Pittsburgh PA 15213, phone: 412 2682268, email:\href{mailto:jgroeger@cmu.edu}{jgroeger@cmu.edu} I would like to thank Blair Richards and Christopher Chidzik at PredictIt for sharing the data. I would like to thank James Turnshek and Jonathan Scott for valuable discussions on this topic.}
}
\author{\begin{tabular}{cc} Joachim R. Groeger\end{tabular}}
\date{ \today }

\begin{document}

\maketitle
\begin{abstract}
In this paper I empirically investigate prediction markets for binary options. Advocates of prediction markets have suggested that asset prices are consistent estimators of the ``true" probability of a state of the world being realized. I test whether the market reaches a ``consensus." I find little evidence for convergence in beliefs. I then determine whether an econometrician using data beyond execution prices can leverage this data to estimate the consensus belief. I use an incomplete specification of equilibrium outcomes to derive bounds on beliefs from order submission decisions. Interval estimates of mean beliefs cannot exclude aggregate beliefs equal to 0.5. 
\end{abstract}
%%%%%%%%%%%
%%%%%%%%%%%

%%%%%%%%%%%%%%%%%%%%%%%%%%%%%%%%%%%%%%%%
\section{Introduction}

The promise that markets and prices can act as information aggregators is considered one of the core advantages of a decentralized economic system.\footnote{The idea dates back to \citet{hayek} and was formalized by \citet{hurwicz}.} If prices are able to fulfill this function in markets for goods, then it is likely that markets for information could be used to aggregate beliefs. Recently a number of researchers have endorsed the view that markets for information, or prediction markets, are superior aggregators of information (see \citet{arrow} and \citet{zitzewitz}).\footnote{In addition, arguments in favor of prediction markets have used variations of the notion of the ``wisdom of the crowd". The idea is that eliciting information from a large number of individuals can lead to superior predictions. This idea was first studied by \citet{galton}, who suggested aggregating multiple guesses on the weight of an ox led to an accurate estimate of the truth.} In these markets, agents trade Arrow-Debreu securities that pay out \$1 if a specific state of the world is reached. Since prices for these securities are on the $[0,1]$ the price could be used as an estimator for the true probability that a state of the world will occur. These claims are hard to test outside of laboratory settings or in situations where the stochastic process is stationary and multiple realizations are observed.\footnote{If the process is stationary and multiple realizations are available, this begs the question why it would be necessary to create a market and not simply estimate the probability distribution directly from the data.} For unique situations such as outcomes of elections, no statistical test for information aggregation can be constructed, because only one realization of the underlying stochastic process is observed. In this paper, I focus on testing whether traders reach a ``consensus" belief. Following the spirit of arguments made in \citet{aumann}, if traders with heterogeneous priors are rational and the market is able to aggregate private information, then traders should eventually update and converge on a consensus belief. If traders still ``agree to disagree," this suggests that the market is not encoding the private information of players. 

An additional difficulty of testing information aggregation in the field is the complexity of most mechanisms. The mechanism in my setting is the limit order book, a completely order-driven market with no market-makers. This is the dominant method of organizing trade in financial markets, including, the most famous prediction market, the Iowa Electronic Market. It is not clear a priori, however, that this mechanism will lead to information aggregation. An example will help illustrate the key issue with the limit order book: Consider a market for two Arrow-Debreu securities for states of the world $0$ and $1$.  Each security pays out 1 if the corresponding  state of the world is reached. Purchasing $1$ is equivalent to supplying asset $0$. Trade can occur only if someone is willing to buy $1$ at $p$ and another party is willing to pay $1-p$ for $0$. The market is populated by informed traders and ``noise" traders. Noise traders are not profit maximizers.  All informed traders agree on the likelihood of the outcome and believe $1$ will occur.  Assume that noise traders are willing to buy position $0$ at all prices above 0.75 (or, equivalently, supply $1$ at all prices below 0.25) and randomly choose a price with equal probability in that range. Informed agents are willing to transact at any price below or equal to 0.25. Execution prices will therefore all be in the interval $(0,0.25]$, falsely suggesting that state $1$ is unlikely to occur. If new informed traders enter later and can observe only execution prices, they will not be able to determine whether the price of 0.25 provides any more information over their private signals. It is then likely that later entrants will rely only on their own private information.

The previous example highlights that the informational properties of limit order market are not obvious and warrant empirical investigation. As mentioned above, determining whether the market has converged on the correct belief is hard to determine in situations where beliefs  for unique outcomes are elicited, such as political events. In addition, simply testing whether the price is above or below 0.5 might also not be sufficient to determine the accuracy of predictions from the market. I thus first consider whether the order book provides clues to traders reaching a consensus belief. If traders all observe the same public history and actions reveal private information, then beliefs should update in the same direction. Recall from the previous example, that trade requires disagreement on the value of the asset. Therefore, as time passes the number of open orders on one side of the market should increase. Instead, I observe large spikes in executions at the end of the market. I then construct portfolios for all traders and measure the probability of each trader shifting from one asset to the another. If traders are reaching a  consensus we would expect traders to be taking similar positions in the market. The data suggest that traders start in different positions and that few traders shift away from their initial positions. Next, I move away from the purely descriptive approach and use more structure to find upper bounds on belief distributions conditional on different points in time in the market. These belief distributions move little as the end of the market approaches. I then place more structure on the decision between submitting a limit order or executing a market order to identify a lower bound for the belief distribution. The goal of this part of the study is to investigate whether an econometrician can act as an aggregator instead of the market. I estimate upper and lower bounds on the belief distribution and estimate intervals for mean beliefs. These intervals have lower bounds below 0.5 and upper bounds above 0.5. This result suggests that market data provides little new information. The most likely explanation for this finding can be found in the earlier example of trade. I identify a significant number of traders who make losses and who execute a number of poorly timed trades. The evidence points to non-rational behavior. These ``noise traders" and the anonymity afforded by the mechanism, allow informed traders to obscure their information. 

The theoretical literature on information aggregation in various strategic situations is very well developed. In the auction setting, \citet{wilson} considers the behavior of the sale price when the number of bidders gets large.\footnote{More recent example of this line of inquiry are \citet{milgrom}, \citet{pesendorfer} and more recently \citet{reny}} \citet{ottaviani2} analyze a simple model of non-strategic trade and discuss convergence of prices to rational expectations equilibria. \citet{grossman} is part of an older literature showing that it is impossible to always guarantee information aggregation. In the context of a prediction market, \citet{manski} shows that prices will not necessarily equal  probabilities. \citet{wolfers} provide a response to \citet{manski}, deriving conditions where market prices are equal to probabilities. 

\citet{ostrovsky} is the most recent theoretical contribution and provides the most general analysis of information aggregation. The author shows that convergence occurs with a finite number of bidders in settings with a market maker and a large number of trading opportunities. The market price will, in finite time, be a consistent estimator of the true state of the world. However, this results does not extend to limit order markets, as the simple examples above show.  In addition, mechanisms for which a position in the market can be ``unravelled", might prevent the aggregation of information, because a trader does not have to commit to a belief and can use information revealed during earlier trade to update their position. In this case, it is not clear that only separating equilibria are possible.

The experimental literature on information aggregation is well developed, starting with the work of \citet{plott} and \citet{plott2}. The mechanisms under investigation in the laboratory are similar to the limit order book. Results suggest that the rational expectations equilibrium provides the best prediction of price behavior. \citet{plott3} discuss the results of a firm's internal prediction market organized around a limit order book and show that prices provide a good predictor of true outcomes. \citet*{shum} continue this investigation at another firm. The market mechanism in that paper more closely resembles a betting market, where trade does not occur. The authors cite the desire to avoid ``bubbles" and coordination problems in their decision to move away from a trading model. \citet{cowgill} consider outcomes from internal prediction markets. Moreover, \citet{zitzewitz} are concerned with investigating the favored long-shot bias more carefully. \citet*{snowberg} and \citet*{forsythe} consider the Iowa Political Stock market for the 1988 Presidential election. All these papers provide evidence that prices can be used as aggregators of information. \citet{rothschild} use individual-level trade from Intrade, a market similar to the platform under investigation in this paper, and are interested in studying the possibility of manipulation of prices. However, the data do not allow the authors to fully re-construct the order book and excludes information on unexecuted limit orders. I am able to completely reconstruct the order book and more importantly, measure trader intentions as well as executions.

%%%%%%%%%%%%%%%%%%%%%%%%%%%%%%%%%%%%%%%%
\section{The Market Institution and Trading Rules}
The data are from the platform ``PredictIt".\footnote{PredictIt is an educational project of Victoria University, Wellington of New Zealand, a not-for-profit university, with support provided by Aristotle International, Inc., a U.S. provider of processing and verification services. The market involves real money and therefore traders are incentivized to provide accurate forecasts.} The data consist of two prediction markets for political events. I first provide descriptive analyses of trading behavior in this market. The market is organized around an electronic limit order book and the market is completely order-driven. An investor trades by submitting limit orders. The minimum tick size is \$0.01. A trader who wants immediate execution must place the order at the best price level on the opposite side of the book, I call these orders ``market orders". An investor who wants to buy or sell more shares than what is currently outstanding at the best price level must submit separate orders for each price level. If a limit order executes against a smaller order, the unfilled portion stays on the book as a new order. Time and price priority between limit orders is enforced. For example, if a trader submits a buy order at a price level that already has other buy orders outstanding, all the old orders must execute before the new order.

Trade occurs in two ways: the first is that the trader encounters another trader who holds the desired asset and is willing to sell. The other is to encounter a trader who is willing to purchase an asset on the opposite state of the world. In other words, buying an option for state $1$ requires a counter-party to purchase an option for the complementary state of the world, $0$, at the desired price. In particular, if the buyer of the $1$ asset is willing to pay $p$ then the trader on the opposite side must be willing to purchase the asset $0$ at price $1-p$.  The platform also creates linked securities for events that have more than two outcomes, such as three political candidates vying for the same position. The platform creates ``yes'' and ``no" securities for each candidate. In these situations, a trader can pursue a more sophisticated strategy. I describe the payment scheme for this setting in the appendix.
%%%%%%%%%%%%%%%%%%%%%%%%%%%%%%%%%

\section{Market Data}
I look at two markets for political events. The first is the market for the Republican Iowa Caucus, the second is on the Supreme Court decision on marriage equality. The latter market had very low daily volumes relative to those of the Iowa Caucus. I therefore focus attention on the Iowa Caucus and provide information on the marriage equality market only in the appendix. PredictIt created markets for each candidate in the Iowa Caucus. There were 14 candidates in this caucus. Donald Trump, Ted Cruz and Marco Rubio were the favorites. These three assets were the most traded by volume. I have data on all orders, trades, modifications and identifiers of traders for all markets. I provide summary statistics for the top three assets in Table \ref{sum_stat}. The average trade price for Cruz (the eventual winner) was 0.45. Rubio options were executing at prices close to 0.07 on average. Trump's average price was 0.62. These prices are closely aligned with the results from various polls.\footnote{See for example polling data at Real Clear Politics \underline{\href{https://tinyurl.com/hl9btyr}{Link}}. Accessed July, 7, 2016.} The market attracted a total of 4452 unique traders. The daily number of active traders was between 11 to 30. However, this number increased as resolution approached. Volumes in these markets were also much higher with daily volume of trades on average between 858 to 1195 securities.

%%%%%%%%%%%%%%%%%%%%%%%%%%%%%%%%%
\input{tables/summary}
\clearpage
%%%%%%%%%%%%%%%%%%%%%%%%%%%%%%%%%
\vspace{.1in}
\noindent
\emph{Trader Profits}

\noindent I now provide some summary statistics on profits for each trader. I break down profits into trading profits, i.e. from buying an asset and selling at a higher price, and the profits from holding on to an asset until the resolution of uncertainty. We also characterize traders as ``Day Traders" (DT) and ``Non-Day Traders". ``Day traders" in financial markets follow a strategy of buying and selling financial instruments within the same day in order to have all positions closed at the end of the day. In my setting since the market operate continuously, a ``day trader" exits the market with zero holdings of any assets in the market.

\begin{figure}[h]\centering\caption{Empirical Distribution of Profits of ``Day Traders" versus ``Algorithmic Trader", Iowa Caucus Market\label{algo_day}}
\includegraphics[width=.75\textwidth]{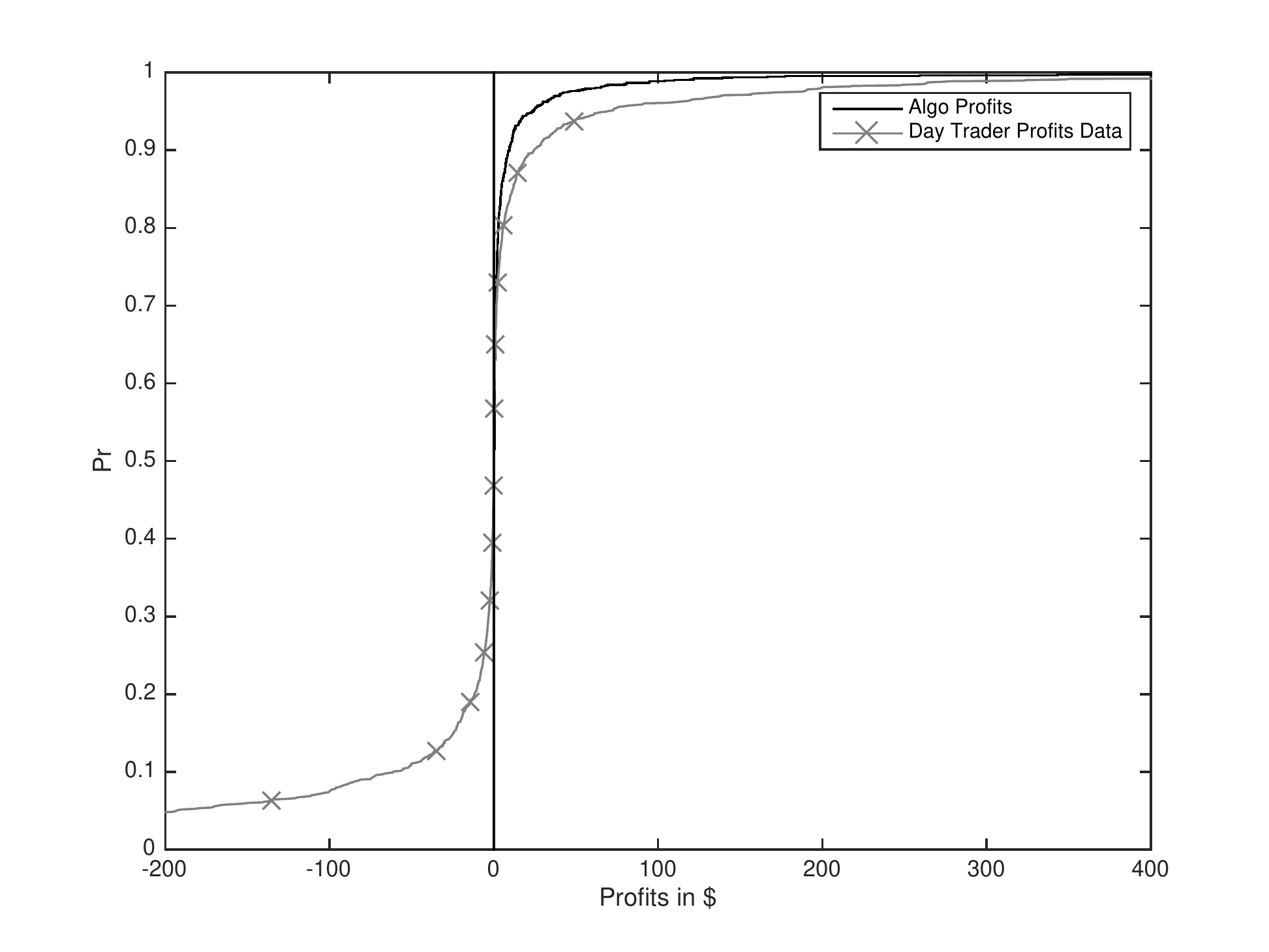}
\end{figure}

Losses are on average larger than profits- less than 1\% of traders make profits greater than \$400; however, about 5\% of traders make losses greater than \$400.
%%%%%%%%%%%%%%%%%%%%%%%%%%%%%%%%
\input{tables/profits}
%%%%%%%%%%%%%%%%%%%%%%%%%%%%%%%%%
The magnitude of losses of day traders is surprising. In order to assess the significance of these losses I use a simple trading algorithm in order to provide a benchmark. I consider the simplest choice rule that is easy to implement in the computer. I use the same buying decisions of the day trader and wait for profitable market orders. I avoid solving an optimal stopping problem and simply wait for the next limit order submitted that could generate positive profits large enough to sell a trader's position. The results are shown in Figure \ref{algo_day}. It is clear that I am able to avoid negative profits by definition and thus outperform the day traders in my data. The profits I net are lower than the day traders who make positive profits, this is mostly due to my reliance on executing as soon as a price arrives that makes a positive profit and not solving for an optimal stopping threshold.

%%%%%%%%%%%%%%%%%%%%%%%%%%%%%%%%%
\begin{figure}[h]\centering\caption{Empirical Distribution of Profits of ``Day Traders", Iowa Caucus Market}
\includegraphics[width=.75\textwidth]{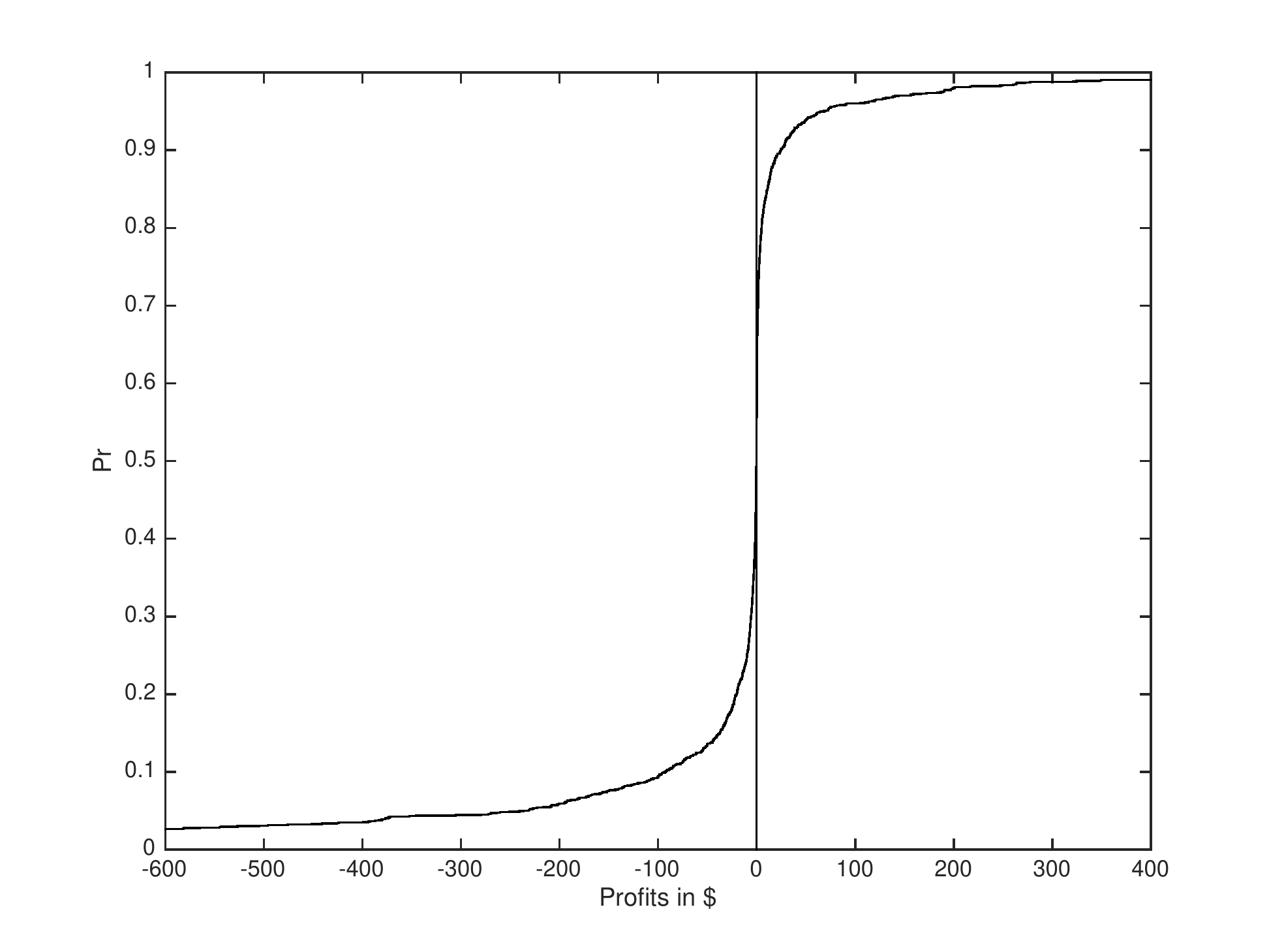}
\centering\caption{Empirical Distribution of Profits of ``Day Traders"  Marriage Equality Market}
\includegraphics[width=.75\textwidth]{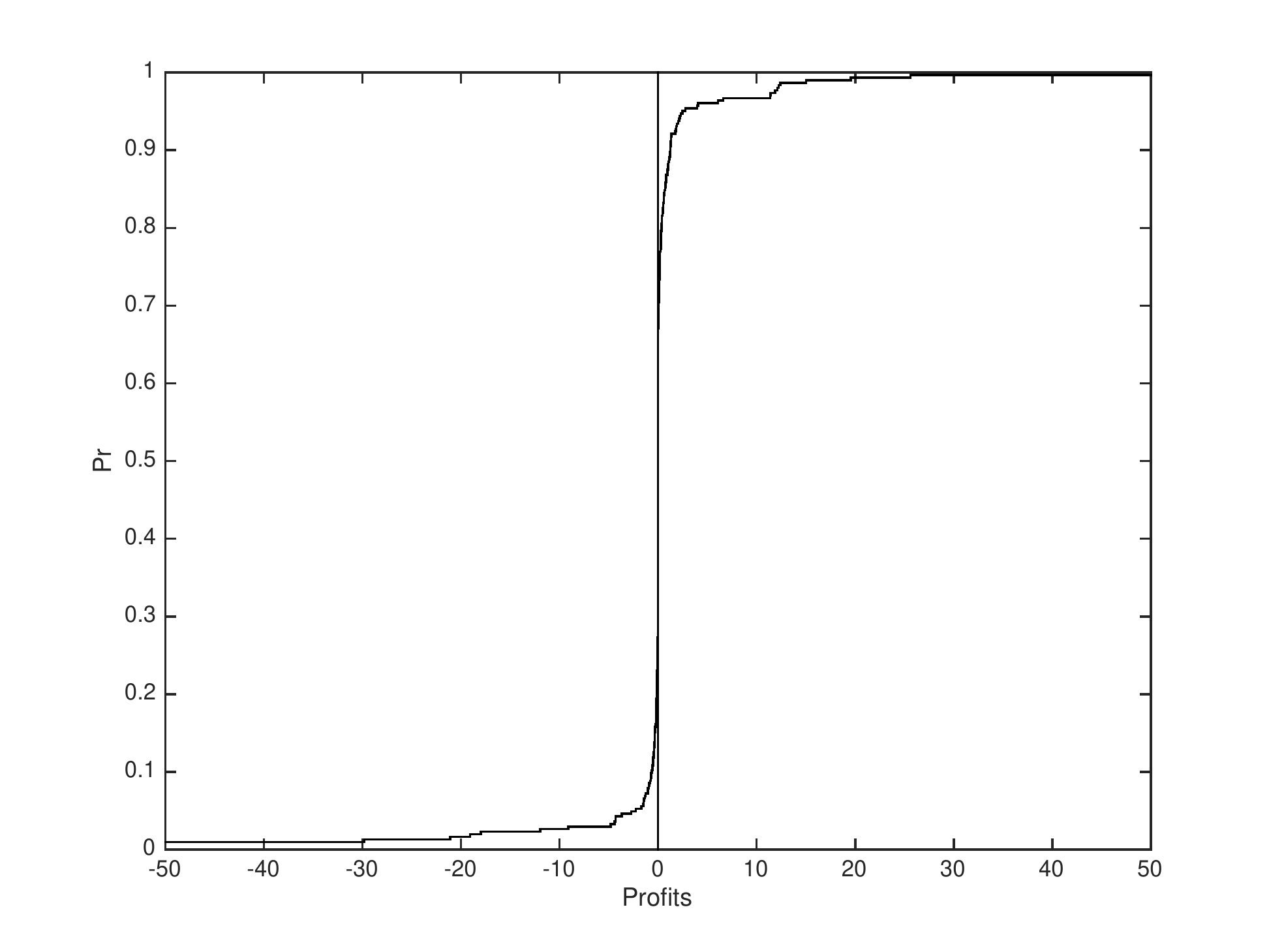}
\end{figure}
%\clearpage
%%%%%%%%%%%%%%%%%%%%%%%%%%%%%%%%%
Together this suggests that a number of ``day traders" have to be treated as ``noise traders" who have no information on events. The existence of poorly performing noise traders raises concerns about whether prices can aggregate information effectively. If noise traders do not set prices that reflect underlying beliefs, informed traders can take advantage of their knowledge and make large profits.

%%%%%%%%%%%%%%%%%%%%%%%%%%%%%%%%%%
%%%%%%%%%%%%%%%%%%%%%%%%%%%%%%%%%

%%%%%%%%%%%%%%%%%%%%%%%%%%%%%%%%%

\clearpage
%%%%%%%%%%%%%%%%%%%%%%%%%%%%%%%%%

\section{Descriptive Evidence on Consensus}
Information aggregation usually involves studying the behavior of prices as the number of players or trading periods grows large. Given the fixed number of players here, I will first focus on the limit when the number of trading times increases, as in \citet{ostrovsky}. Price paths should therefore converge to the ``true probabilities" as we approach the resolution of uncertainty.  In addition, as time increases, if information is revealed through price paths or bid and ask paths, we should observe traders using market information to update their own private beliefs. Updating of beliefs is hard to directly observe, though. I identify dimensions in the data that might provide an indication of whether information is encoded in some form in the market. These include studying the profitability of traders against their first entry time, and most importantly, the portfolio holdings of each trader across time.

\begin{figure}[h]
\centering\caption{Execution Prices Iowa Caucus\label{avg_price_io}}
\includegraphics[width=.75\textwidth]{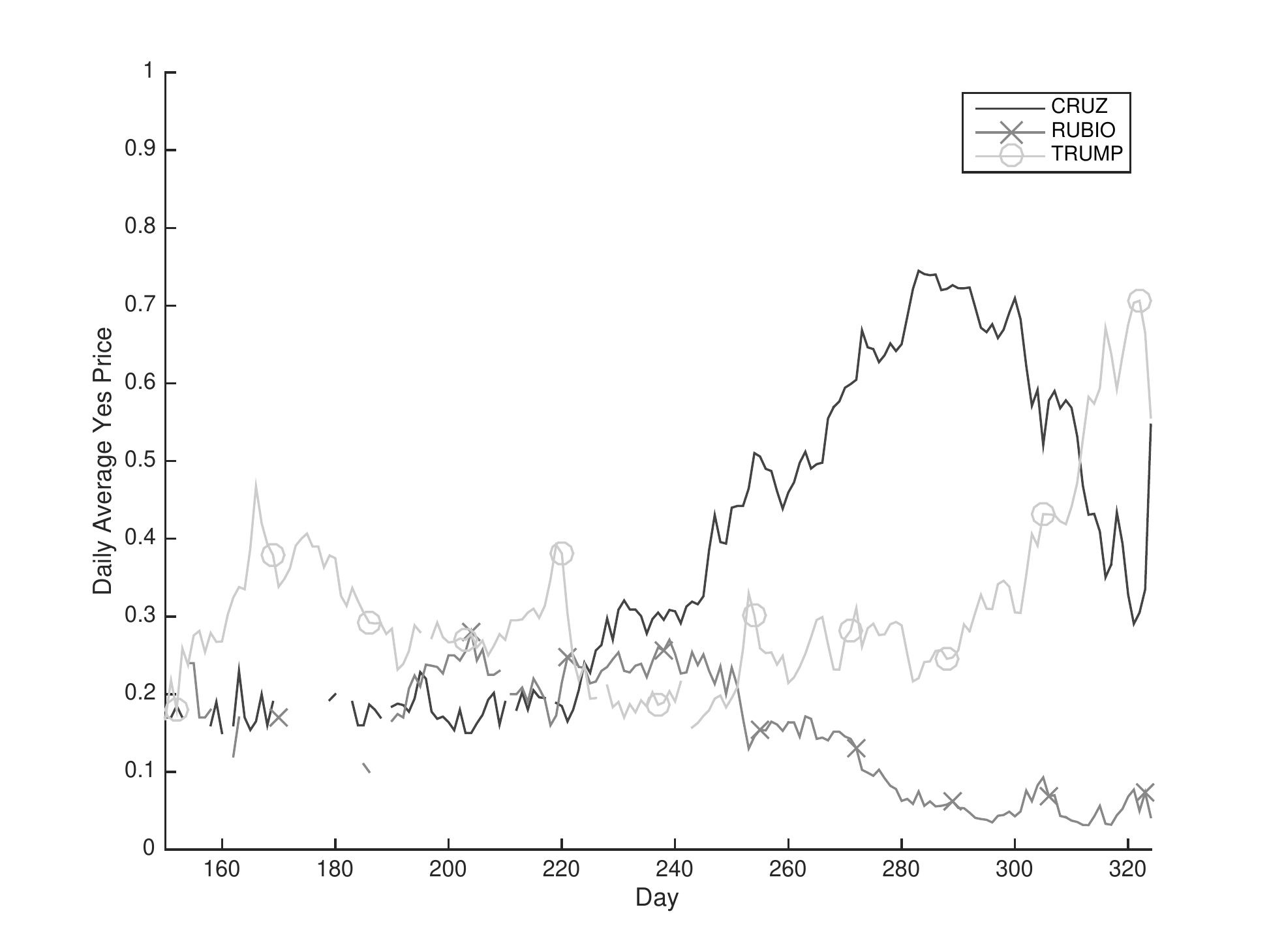}
\end{figure}

\begin{figure}[h]
\centering\caption{Daily Volume Iowa Caucus\label{avg_vol_io}}
\includegraphics[width=.75\textwidth]{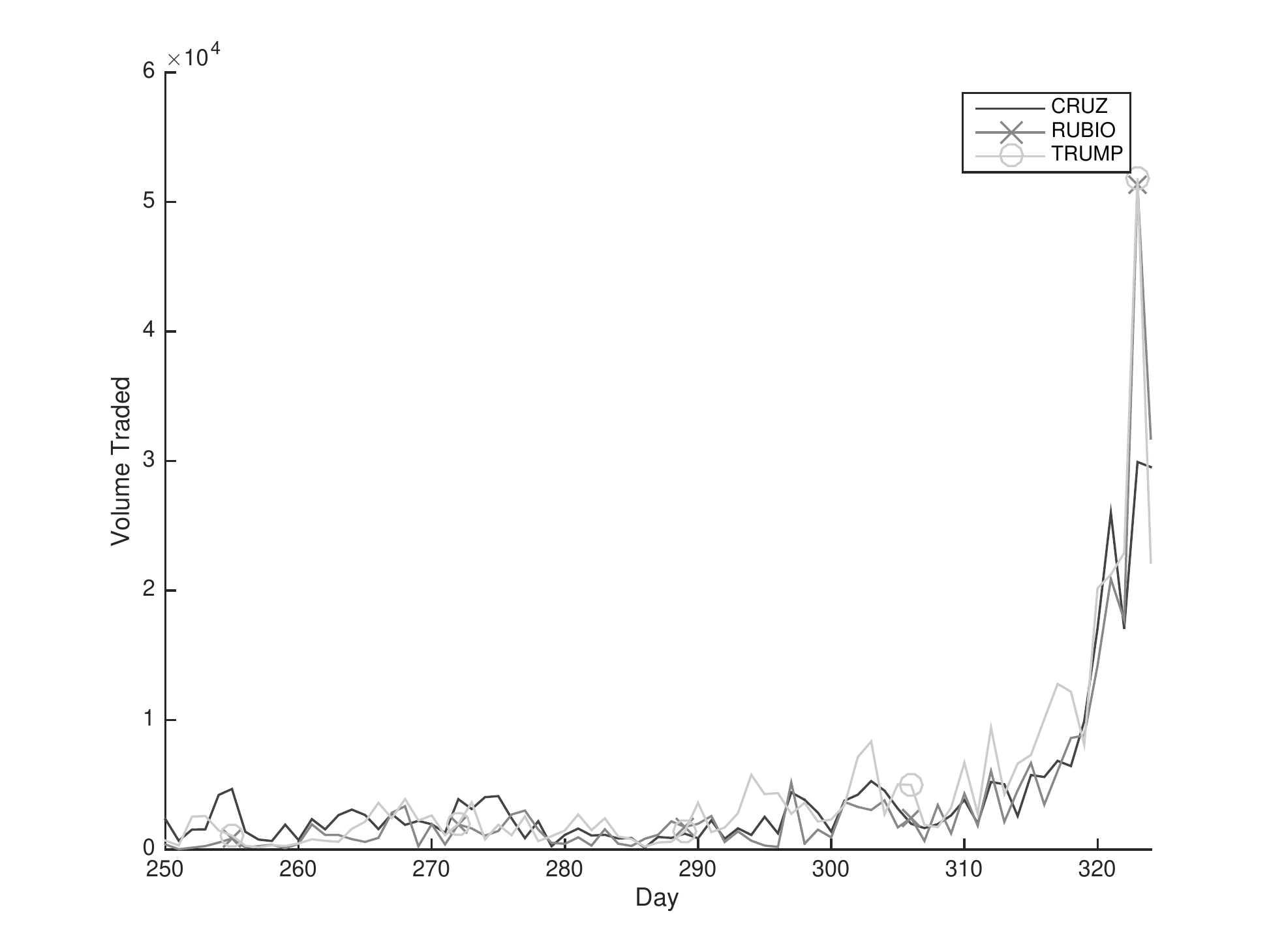}
\end{figure}

In Figures \ref{avg_price_io} and \ref{avg_vol_io}, I show volumes and prices for the top three traded securities. Securities for Cruz traded on average at prices below 0.5 for at least 250 days. There was a steady increase until day 280. The price then started decreasing, hitting a low of 0.3 close to the end of the market. Trump assets followed a price path around 0.4. Similarly, Rubio was actively traded at prices around 0.2. 
%%%%%%%%%%%%%%%%%%%%%%%%%%%%%%%%%
%%%%%%%%%%%%%%%%%%%%%%%%%%%%%%%%%
The volume of trade spiked as the market reached its conclusion. This is the first piece of evidence against a consensus. In particular, we would expect to see a drop in volume as traders' beliefs converge and there is little disagreement between traders. As disagreement is reduced, more traders will switch to one side of the market and would have to rely only on noise traders to generate trading opportunities. The spike in execution volume makes this claim harder to support.  I also look at the volume of open orders in the book. If a consensus has been reached and traders are awaiting the arrival of noise traders, then most open order volume should be collecting more on one side of the market. 
%%%%%%%%%%%%%%%%%%%%%%%%%%%%%%%%%

\begin{figure}[h]
\centering\caption{Volume of Open Orders Across Time, Cruz, Iowa Caucus\label{open_vol_cruz}}
\includegraphics[width=.75\textwidth]{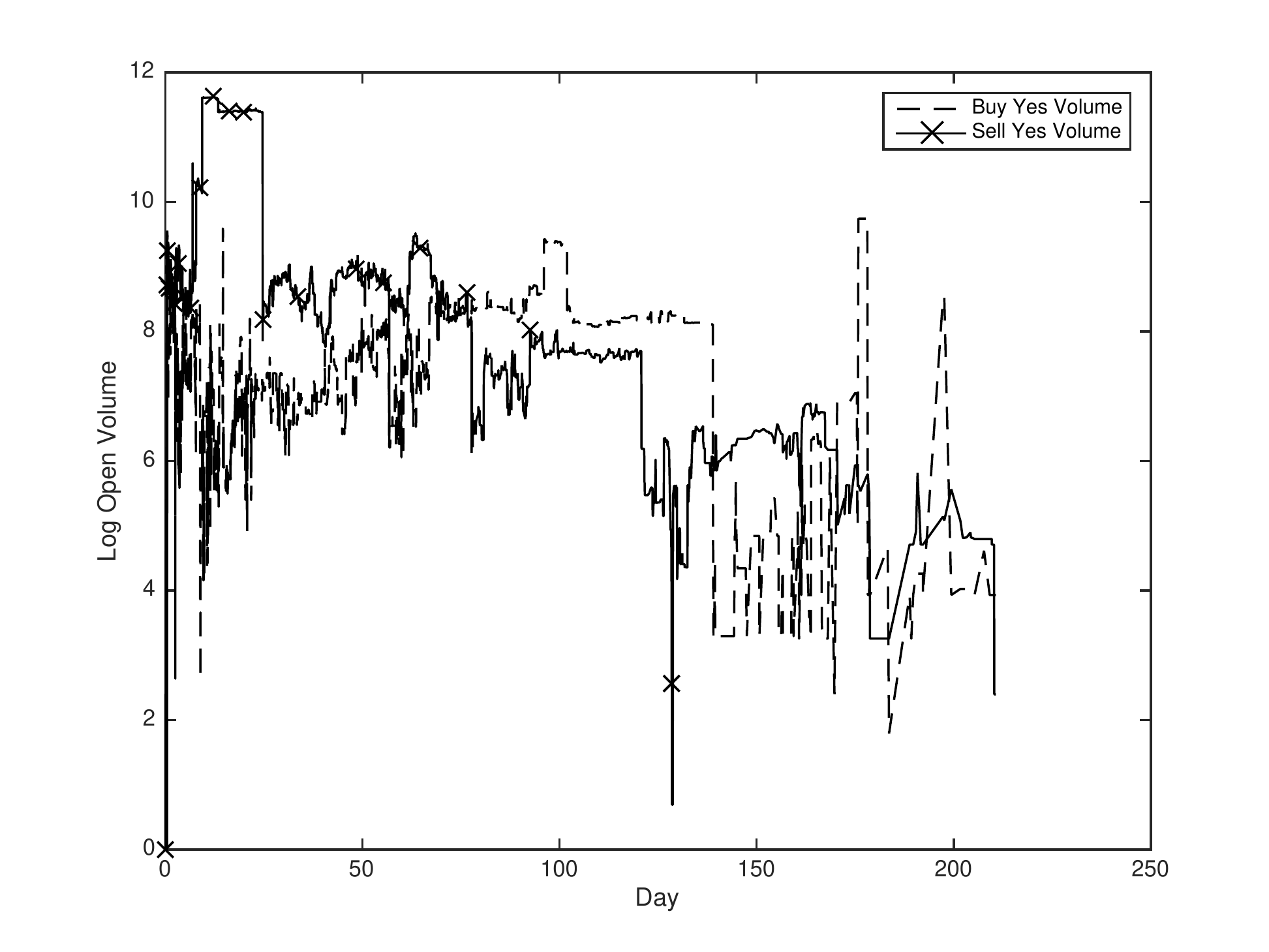}
\end{figure}

\begin{figure}[h]
\centering\caption{Volume of Open Orders Across Time, Trump, Iowa Caucus\label{open_vol_trump}}
\includegraphics[width=.75\textwidth]{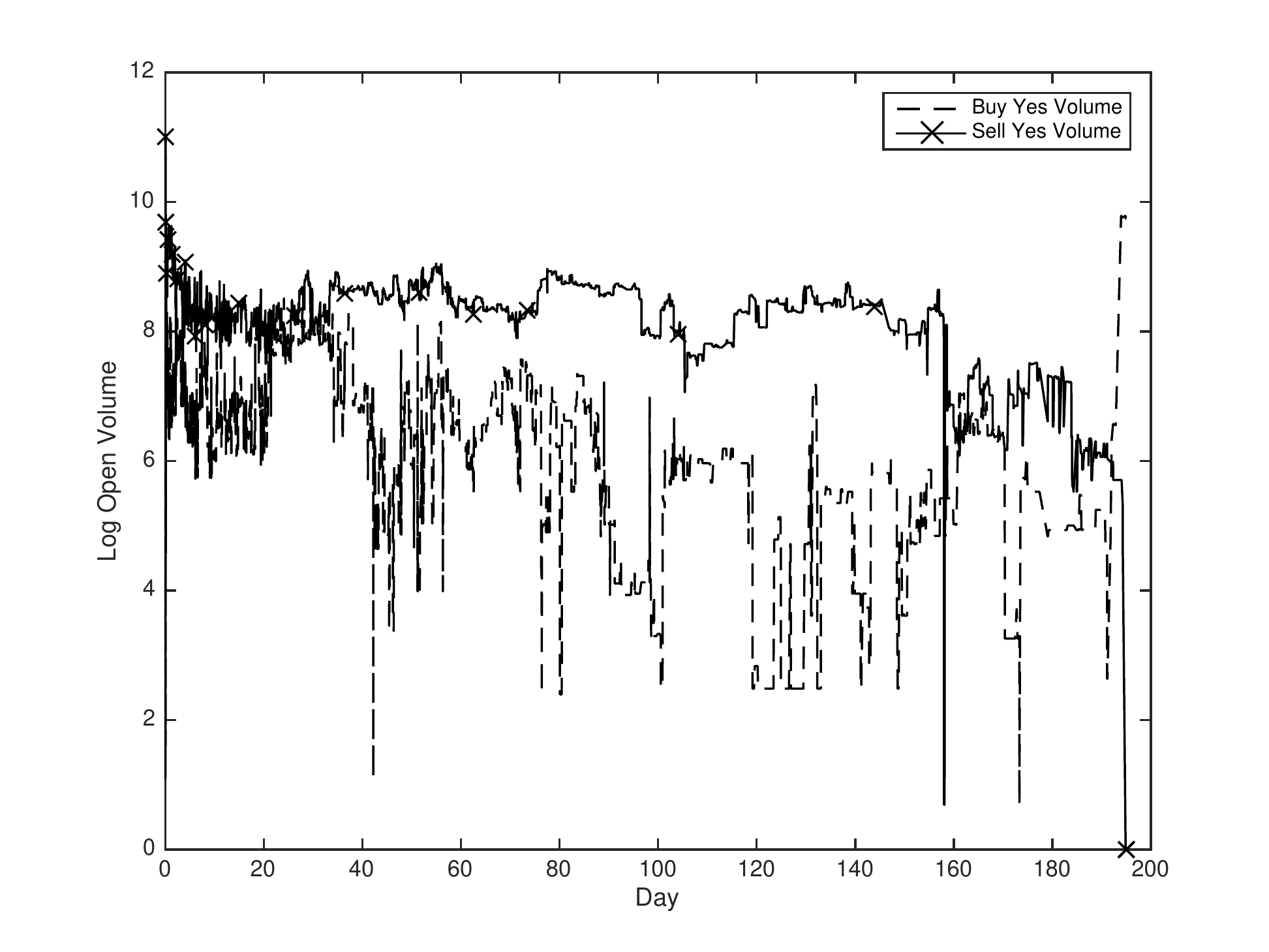}
\end{figure}

\begin{figure}[h]
\centering\caption{Volume of Open Orders Across Time, Rubio, Iowa Caucus\label{open_vol_rubio}}
\includegraphics[width=.75\textwidth]{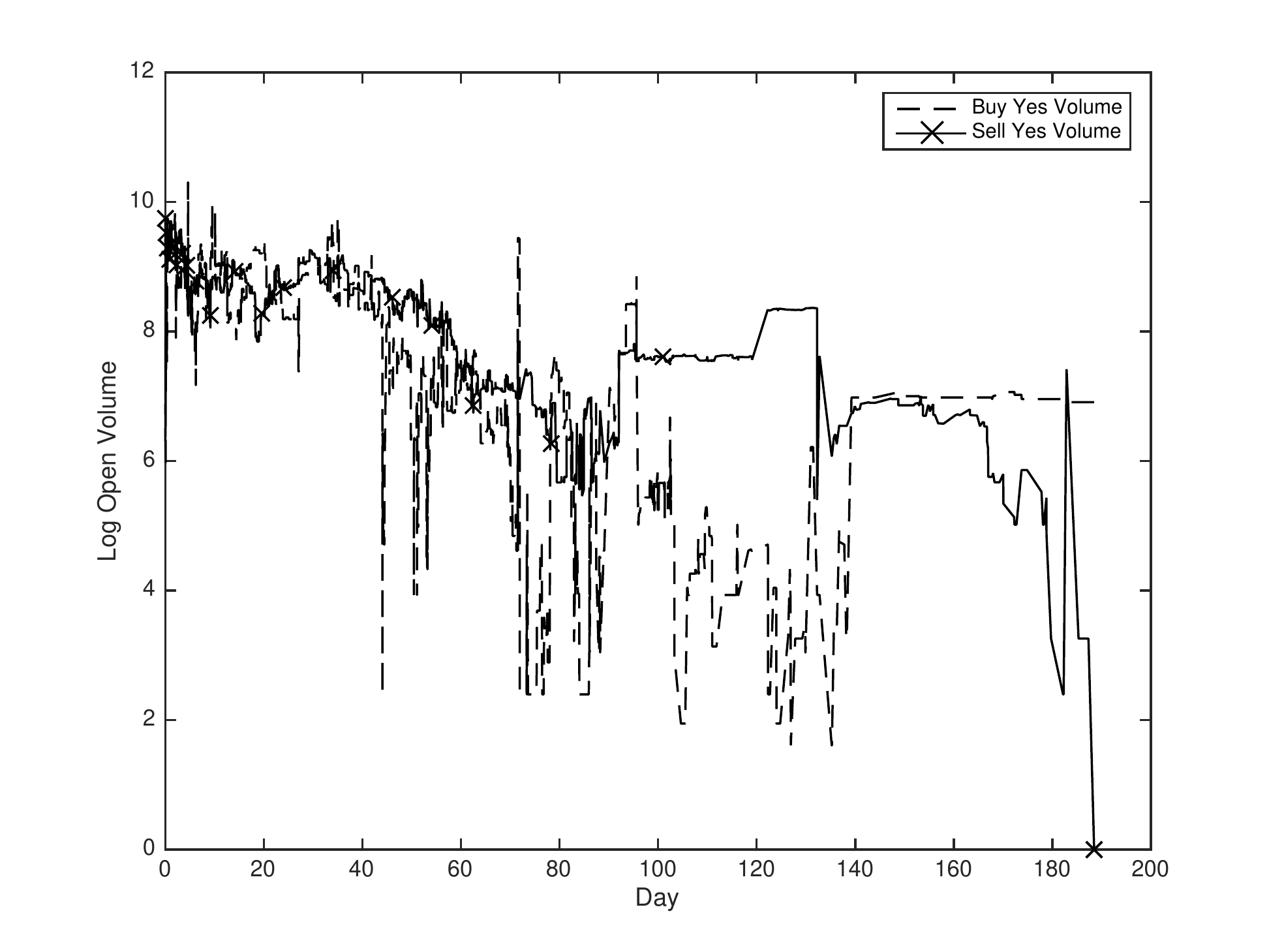}
\end{figure}

In Figures \ref{open_vol_cruz}, \ref{open_vol_trump}, and  \ref{open_vol_rubio}  I show the log quantity of open buy and sell orders for each of the top three candidates ``yes" assets. There is no clear increase in either one of the markets. There are no noticeable increases in open volumes over time, again providing evidence against a convergence to consensus.

\clearpage
%%%%%%%%%%%%%%%%%%%%%%%%%%%%%%%%%

\vspace{.1in}
\noindent
\emph{Portfolio Changes Over Time}

\begin{figure}[h]
\centering\caption{Portfolio Shifts, Iowa Caucus, 20 days prior to End of Market\label{port_shift_io}}
\includegraphics[width=.75\textwidth]{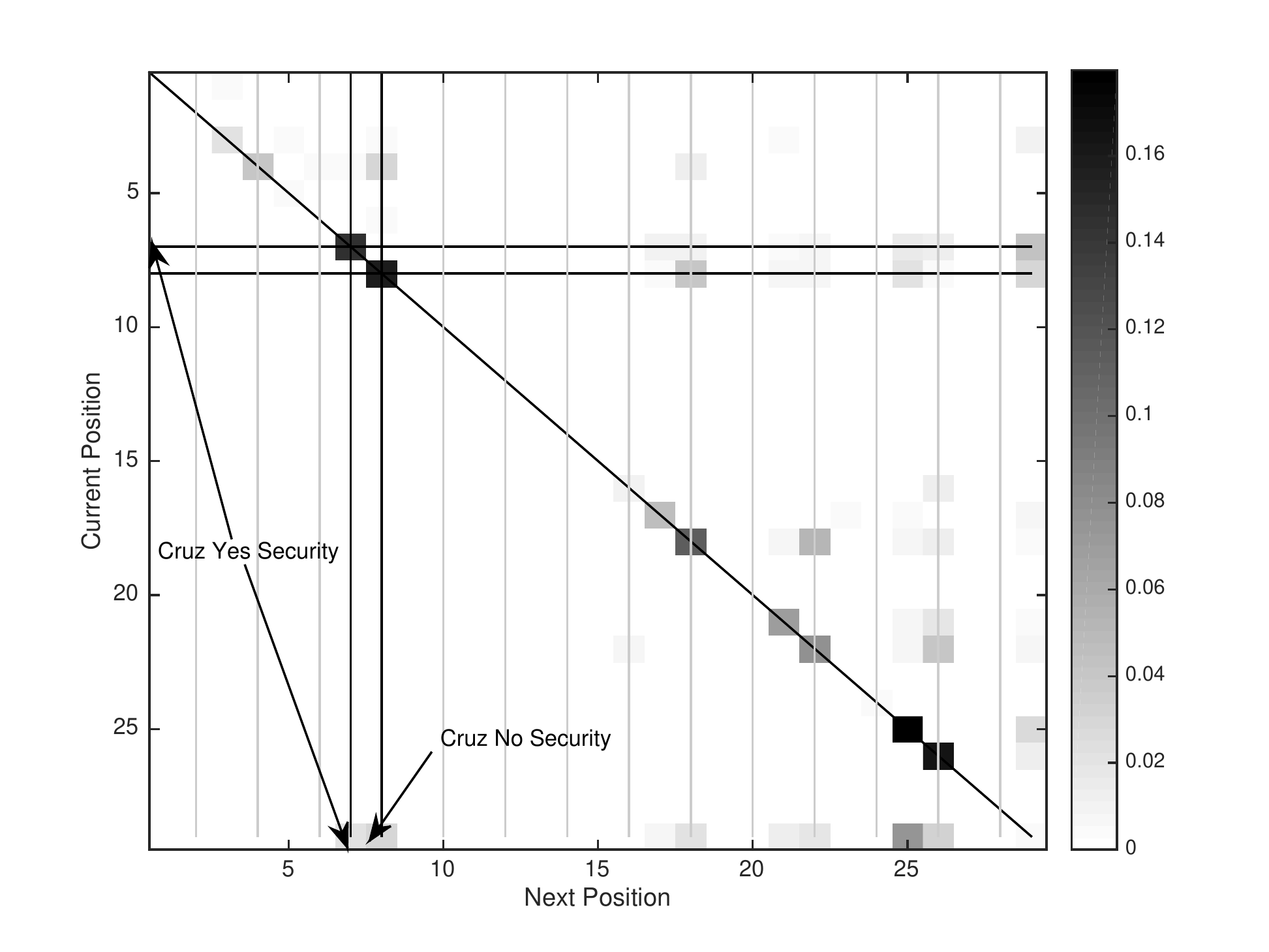}
\end{figure}

\begin{figure}[h]
\centering\caption{Portfolio Shifts, Iowa Caucus, 10 days prior to End of Market\label{port_shift_io_2}}
\includegraphics[width=.75\textwidth]{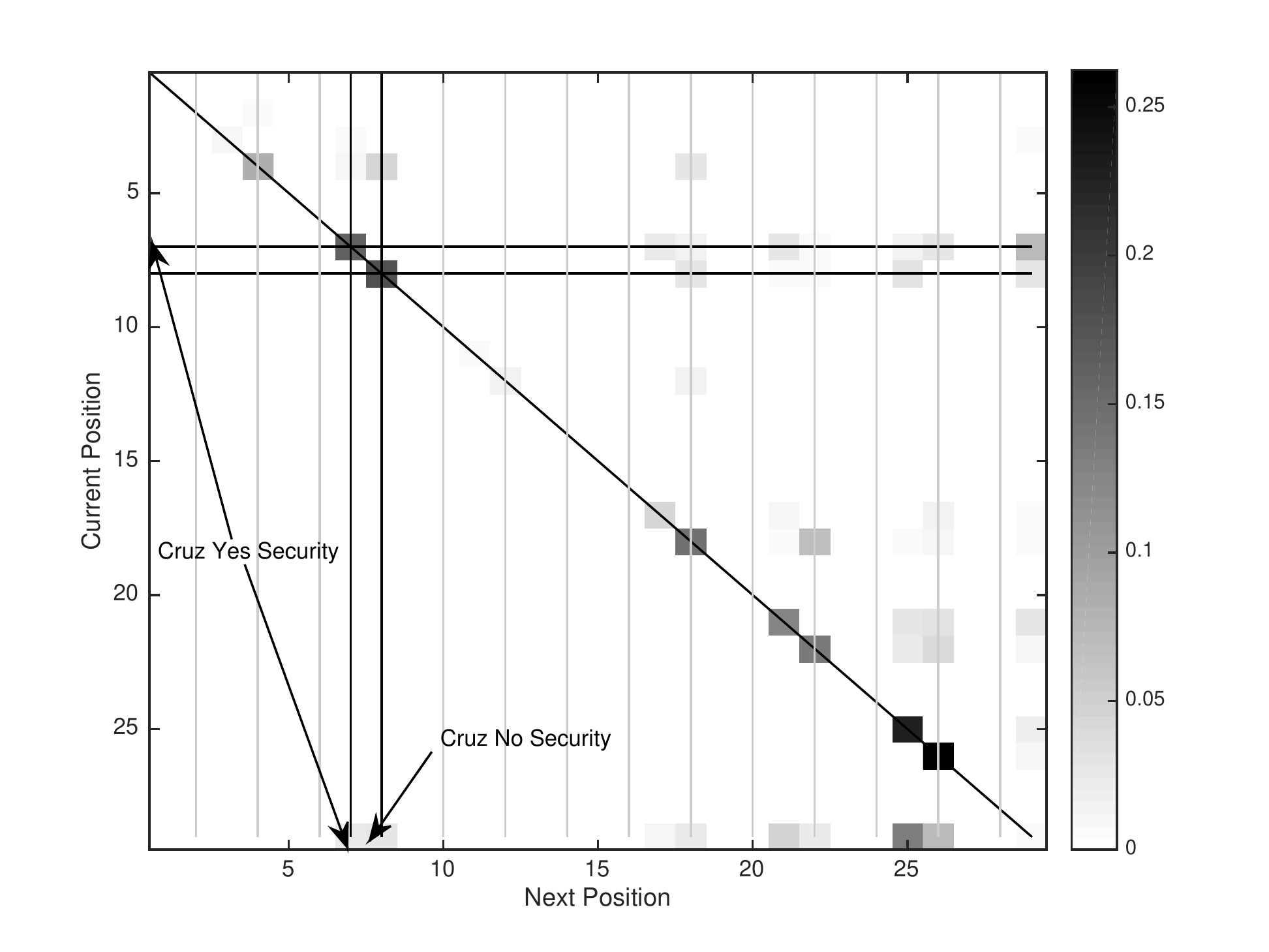}
\end{figure}

\noindent
I now construct portfolio holdings for each trader. If a trader realizes he is on the ``wrong" side of the consensus, he will try and shift his portfolio away from his current position. If the rest of the market is doing the same, then it is likely that there will be fewer orders executed, as most traders will be seeking to unravel their positions. I show the ``transition probability" of the portfolio moving from one asset to another. As consensus is approached, we should see the majority of shifts to the same state of the world. Figures \ref{port_shift_io} and \ref{port_shift_io_2} are visualizations of the ``average transition matrix" for portfolio holdings. Specifically, I estimate a transition matrix of portfolio positions for each trader and average across traders to generate the final transition matrix. The y-axis shows the current asset position and the x-axis shows the next asset position taken after a successful execution. Entries on the downward sloping diagonal represent traders purchasing more securities on their current position. Off diagonal entries are shifts to different securities. There are 28 positions a trader could potentially take in this market. Darker squares represent higher probability entries. Most entries are on the diagonal, i.e., traders purchasing more securities they already hold. If consensus was approaching, traders should be shifting from different ``y-positions" to similar ``x-positions". Yet, there is little evidence of this. Cruz is the eventual winner and many traders- even close to the end of the market- are still trading against him winning, which can be seen in Figure \ref{port_shift_io_2}.
%I have highlighted the Cruz ``yes" security on the x-axis. If information is aggregated and traders can learn as the market progresses and time passes we would expect to see many entries along the vertical line shown for the Cruz asset. In addition, traders might not want to purchase assets on Cruz winning, however, they might purchase securities against other candidates winning. These assets are highlighted by vertical grey lines. Again, most activity is on expanding existing positions and not shifting away from potentially ``bad positions". Interestingly, as we approach the end of the market, traders are not shifting portfolios to the correct outcome and some are indeed shifting deeper into positions against Cruz winning. 
%%%%%%%%%%%%%%%%%%%%%%%%%%%%%%%%%
\clearpage

\begin{figure}[h]
\centering\caption{Unexecuted Buy Orders Relative to Last Position, Iowa Caucus\label{open_buy_io}}
\includegraphics[width=.75\textwidth]{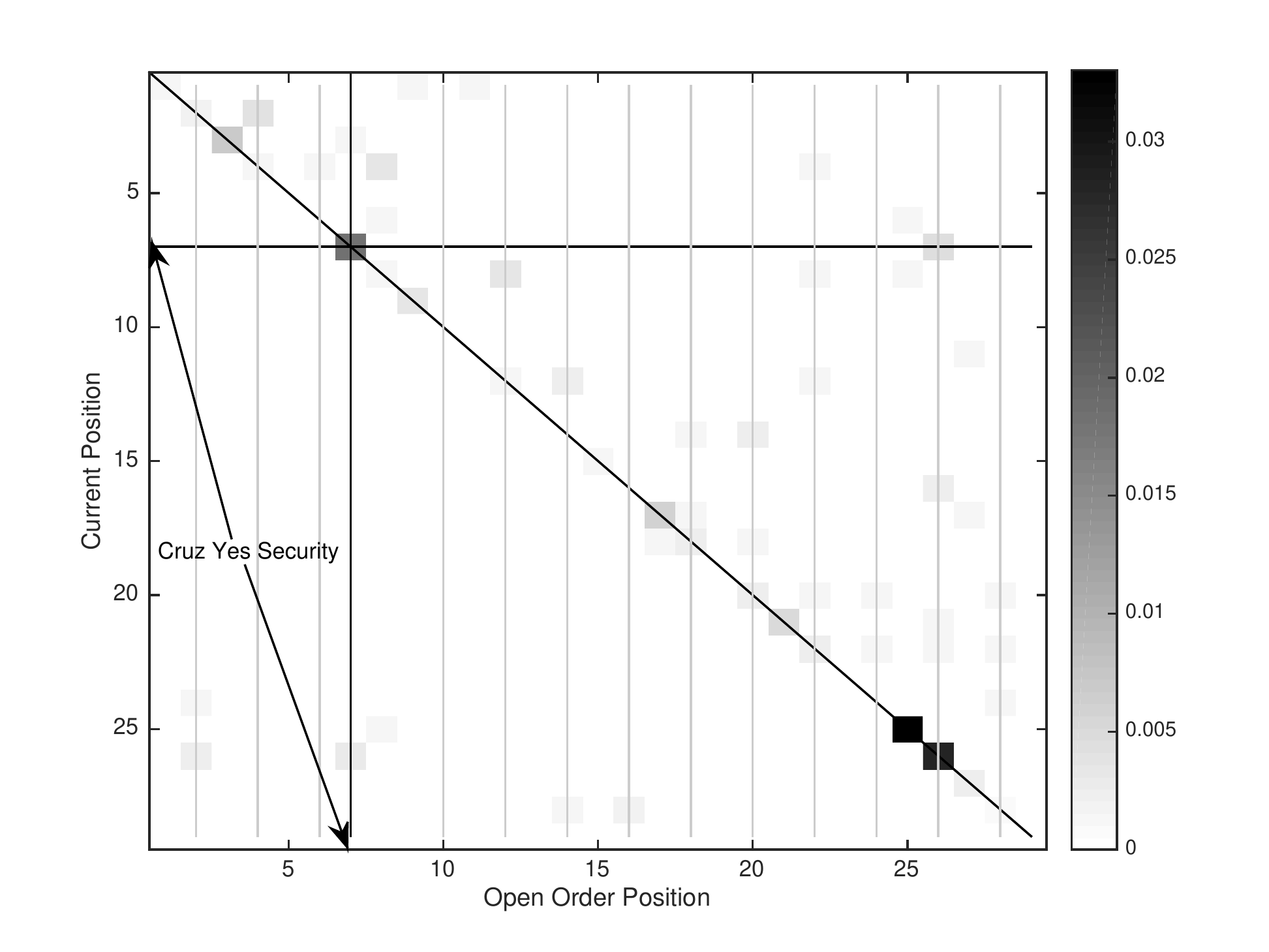}
\centering\caption{Unexecuted Sell Orders Relative to Last Position, Iowa Caucus\label{open_sell_io}}
\includegraphics[width=.75\textwidth]{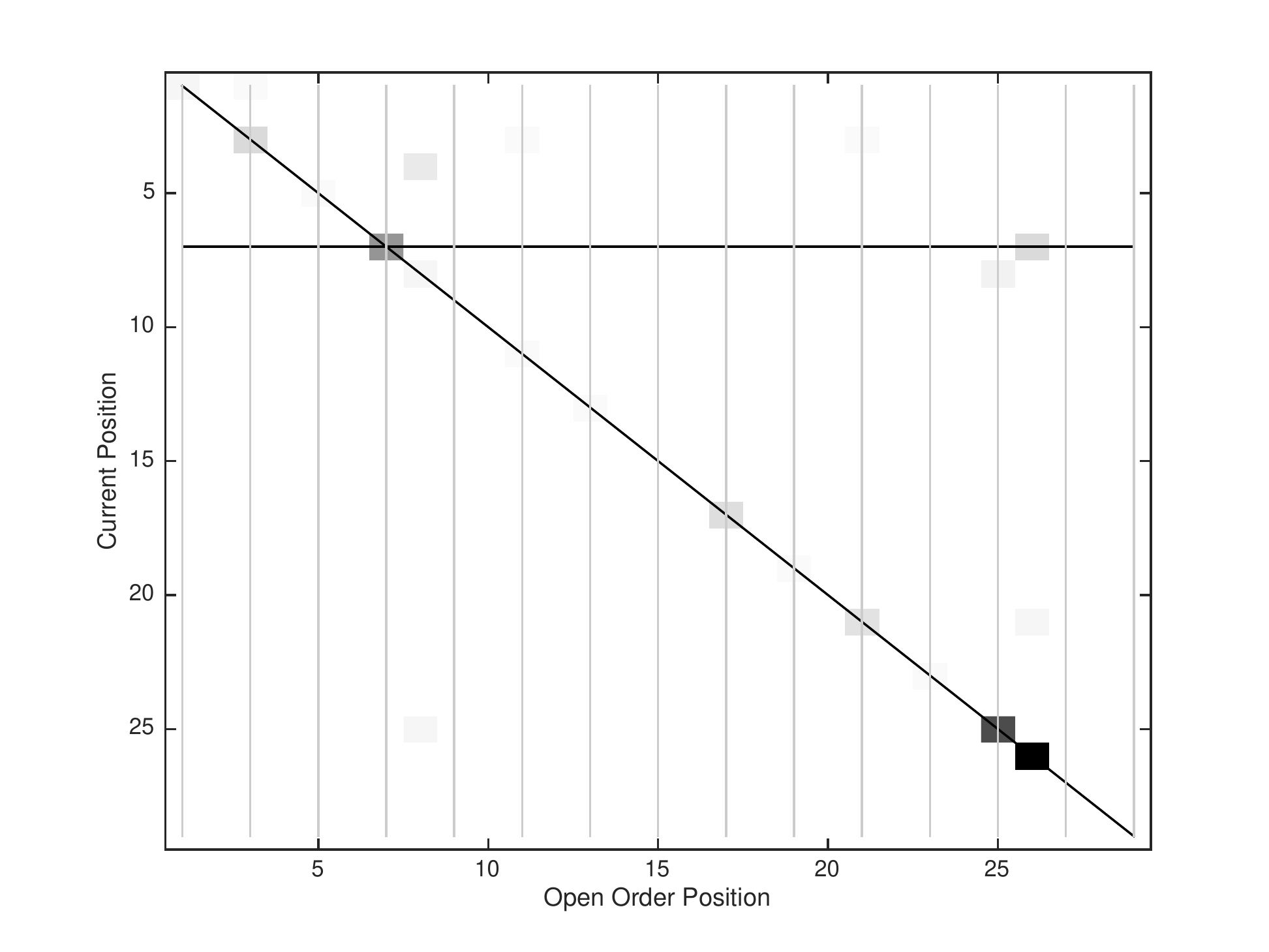}
\end{figure}

%%%%%%%%%%%%%%%%%%%%%%%%%%%%%%%%%
In the appendix, I show the same analysis for the top 30 traders as measured by their prediction profits. The patterns are very similar to the above. I now look at order positions. Traders might not be able to execute their portfolio shifts; their orders, however, reveal their desired position. In Figures \ref{open_buy_io} and \ref{open_sell_io} I show the open buys and sells. A number of traders want to shift into buying Cruz ``yes" assets and a number who would like to shift into ``no" assets for other candidates- this is shown by the squares in the south-east corner of the picture. However, there is no clear systematic pattern suggesting traders moving to the same position. 
%%%%%%%%%%%%%%%%%%%%%%%%%%%%%%%%%
\clearpage
%%%%%%%%%%%%%%%%%%%%%%%%%%%%%%%%%

%%%%%%%%%%%%%%%%%%%%%%%%%%%%%%%%%

%%%%%%%%%%%%%%%%%%%%%%%%%%%%%%%%%

%%%%%%%%%%%%%%%%%%%%%%%%%%%%%%%%%
\clearpage

%%%%%%%%%%%%%%%%%%%%%%%%%%%%%%%%%

%%%%%%%%%%%%%%%%%%%%%%%%%%%%%%%%%
\subsection{Profits by Entry Time}
\begin{figure}[h]
\centering\caption{Empirical Distribution of Profits by Entry Time, Iowa Caucus\label{pi_time_io}}
\includegraphics[width=.75\textwidth]{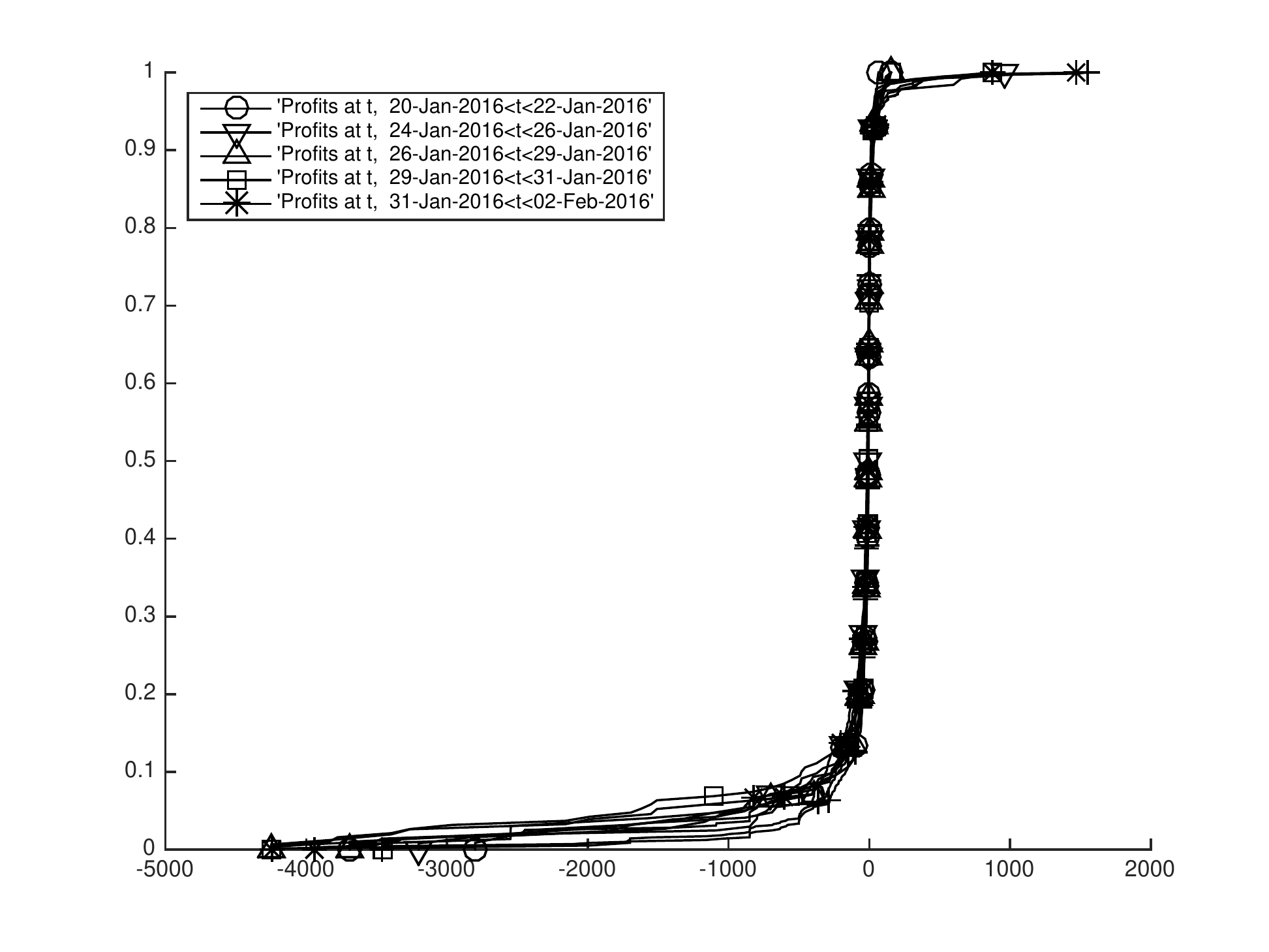}
\end{figure}

I return to the description of profits for traders with positive terminal holdings and the relationship between profits and entry time. If traders enter later and can use the history of the visible order book to update their beliefs on terminal outcomes, then we would expect these players to make systematically different profits. We show the empirical distribution, starting from about six months before the end of the market going all the way to 20 days before the end of the market. The results are shown in Figure\ \ref{pi_time_io}. It is hard to detect visual differences between the distributions. I run a K-sample Kolomogorov-Smirnov Test and find the distributions not to be statistically different at 95\% confidence.\footnote{The test statistic is taken from \citet{kiefer} and is the supremum of the set of distances between each sample distribution function and the distribution function obtained from pooling all observations. The limiting distribution of this statistic is derived in \citet{kiefer}. However, this will not be valid in our setting, given the dependence of the samples. I therefore opt to use a subsampling procedure to characterize the convergence rate and to estimate critical values. \citet{politis} provide details on the consistency of this approach, and an overview is in the appendix.} 
%%%%%%%%%%%%%%%%%%%%%%%%%%%%%%%%%
\clearpage
%%%%%%%%%%%%%%%%%%%%%%%%%%%%%%%%%
\subsection{Summary of Descriptive Analysis}
I find a significant number of noise traders who seem to pursue suboptimal strategies, i.e., making negative profits. Portfolio holdings suggest that non-noise traders enter the market with a fixed belief that is not updated. Traders are anonymous, and there is no way to identify a noise trader without observing her private order and trade history. As a result, updating beliefs based on the observable part of the book leads to biased beliefs. An informed trader can obscure information in the presence of noise traders. 

The platform succeeds in attracting a large number of traders, and it is possible that their actions might partially reveal private information. This seems likely, given that traders seem to ignore public information. I next investigate whether an econometrician can act as a surrogate information aggregator and use order submission decisions to estimate consensus beliefs. 

%%%%%%%%%%%%%%%%%%%%%%%%%%%%%%%%%
\section{Identifying Consensus with an Incomplete Model}
In this section, I use an incomplete model to determine whether the previous descriptive findings still hold with more structure. The model will allow me to use more information than the transaction prices. In particular, order submissions can be used to infer private information, regardless of whether the order executes or not. Models of trade on limit order markets are known to be quite complex and require a number of simplifying assumptions to make any progress in analyzing equilibrium behavior. See for example \citet{parlour} and \citet*{goettler}. These models simplify the action space or restrict attention to equilibrium refinements. The only structural work I am aware of is \citet*{miller_lom} and \citet*{miller_lom_2} who estimate a model where traders have one opportunity to submit a market or limit order. One major difficulty with modeling limit order markets is that trade occurs ``asynchronously", i.e., a limit order submitted now will execute with a market order that arrives later, potentially conditioned on different information. In other words, if a trader submits an order that reveals some of his private information this will provide an advantage to later players, which affects the probabilities of their own order executing. In addition, the standard organization of limit order books precludes traders from tracking other traders' behavior and the exact timing of orders that are in the book. Therefore, a trader cannot use the current  bid-ask spread to accurately determine when an order was submitted, as an order might become visible to the market either because a new order has arrived within the spread or the spread has widened to include orders that were already present in the book. 

I side-step some of these difficulties by constructing an ``incomplete model" of behavior in this market. This is akin to the approach of \citet{haile_tamer} and \citet{sutton2}. The idea, outlined in detail in \citet{sutton2}, is to construct an ``equilibrium configuration" in outcome space instead of specifying the game and deriving strategies. This idea was then adapted and extended in \citet{haile_tamer} to identify primitives, without having to completely specify the equilibria. This generally leads to sets of inequalities that provide partial identification of parameters of interest.
%\footnote{There is a parallel to the literature on the identification common value auctions. The securities in this market have an unknown value that is the same to all players. In the auction context, the econometrician seeks to identify the signal distribution and the distribution of the common value from only bids. This is what leads to a non-identification result; the econometrician only has one bid and wishes to identify to pieces of information from it. In my setting, the goal is to only identify the regression of the true state of the world conditional on private signals. In the auction context this object is identified.} 
The general setup and notation are outlined below.

\begin{itemize}
\item True state of the world: $\omega_0\in\Omega=\{0,1\}$. Nature privately draws the true state of the world. Let $P\equiv \Pr(\omega_0=1)$ be the true probability.
\item Players are small relative to the market and there is a large number of players. There is a set of informed players $\mathcal{N}_I$ and a set of noise traders $\mathcal{N}_N$ who have no information. Whether a trader is informed or uninformed is determined exogenously by nature. Each potential trader is randomly assigned to each set and this information remains private.\footnote{Because the market is large this setup is different from that of \citet*{engelbrecht}, where the informed bidder knows she is playing against uninformed bidders. In my setting, this is unknown and the number of each type is also unknown. In addition, I assume that the noise traders are unsophisticated and non-optimizing players. This is in line with the data on day trader behavior. }

\item Each trader $i$ in $\mathcal{N}_I$ receives a signal on the underlying true probability and forms belief $q_{it}$ on the probability that $\omega=1$, which is potentially a function of the complete history $h_t$.  Uninformed traders are ``noise" traders who receive no private information and are not profit maximizers. An informed trader should be thought of as an ``insider," either someone who has put effort into learning more about the state of the world or an insider with access to information before others.

\item Entry into the market is stochastic and the identity and types of traders remain unknown to all participants. Entry times are drawn independently and identically for all traders from an exogenous distribution. In addition, further opportunities to trade arrive randomly and are determined by an exogenously given hazard function. 
\item Trade occurs continuously on the time interval $[0,1]$. At $t=1$ all uncertainty is resolved and the true state of the world is revealed.

\item Traders buy and sell Arrow-Debreu securities. A security pays out \$1 if a specific state of the world is reached at a deadline. There are two securities, one for each state of the world. Trade is organized around a limit order book. The action space consists of the decision to make a limit or market order, and in the case of the limit order, the price for the order. The action space is denoted $\mathcal{A}\equiv\Omega_a\times\mathcal{O}\times\mathcal{P}$, where $\Omega_a\equiv\{0,1\}$ is the decision on the state of the world. The order type decision is denoted $o\in\mathcal{O}\equiv\{0,1\}\equiv\{\textrm{Market Order},\textrm{Limit Order}\}$ and the price decision $p_t\in\mathcal{P}\equiv[0,1]$. I assume that the size of an order is exogenously given. Denote by $y_\omega$ the holdings for a trader of asset $\omega$. 
\item The public information set at time $t$ is given by the five best buys and sells at time $t$ for the two securities. The information is summarized in $h_t$, which is defined as:
\[
h_{t}=(\sigma_{0t},\sigma_{1t},\sigma_{0t}^x,\sigma_{1t}^x,h_{t-1})
\]
where $\sigma_{kt}$ is the vector of the five best sells and buys for state $k\in\{0,1\}$ and $\sigma_{kt}^x$ is the vector of quantities associated with each order. Orders are anonymous, and there is no way for traders to make inferences on the identities of their opponents. 
\item The equilibrium probability that a limit order at price $p$ executes is denoted by $\phi_t(p,y)$. Again $\phi_t$ can depend on the entire public history, $h_t$. 

\item An informed trader seeks to maximize their terminal payoffs. Traders are forward-looking to the extent that they select limit prices to maximize their expected profits. They ignore, however, future trading opportunities and the effects of a current order on future states. 
\end{itemize}
%%%%%%%%%%%%%%%%%%
\subsection{Equilibrium Configurations and Bounding Beliefs}
I now provide structure on equilibrium configurations and focus only on the behavior of non-noise traders. I drop the $\omega$ subscripts from the following derivations. I focus on the decision to purchase the asset for state of the world $\omega=1$, i.e., the decision of traders who have belief that $q_{it}>1/2$. Equivalent bounds can be derived for traders on the opposite side of the market and are used in estimation to identify the complete distribution of beliefs.

\begin{figure}[h]
\centering\caption{Upper Bound on Private Information Distribution Ted Cruz Conditional on History, Iowa Caucus\label{ub_hist_4}}
\includegraphics[width=.75\textwidth]{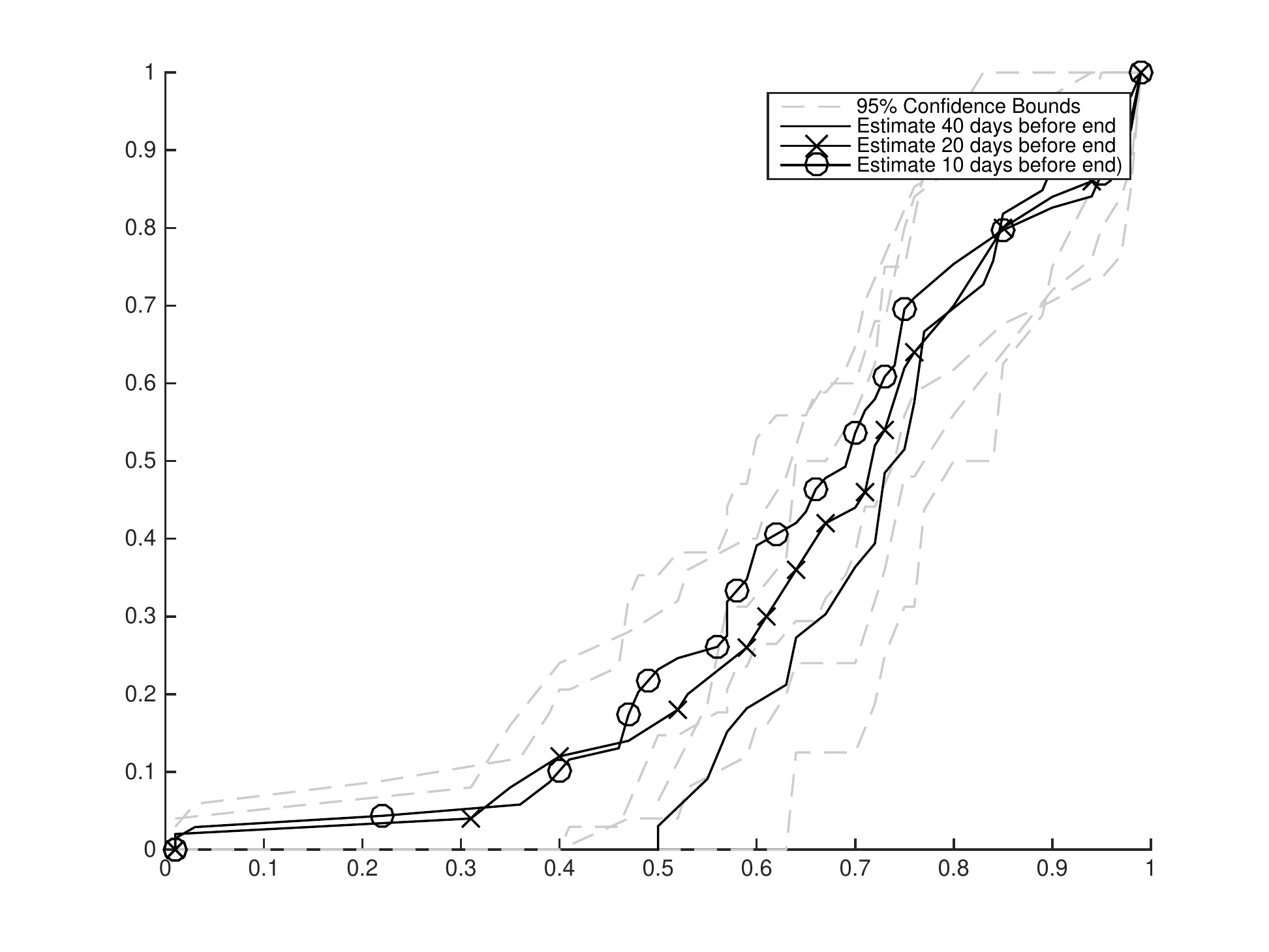}
\end{figure}

%%%%%%%%%%%%%%%%%%
\begin{assumption}\label{pos_prof}
All traders, except noise traders, choose order prices or execute market orders that yield positive expected profits. Specifically, at time $t$, trader $i$ with beliefs $q_{it}$ chooses market or limit price $p_t$ such that
\begin{eqnarray}
q_{it}\geq p_t
\end{eqnarray}
\end{assumption}
This provides an upper bound on the distribution of beliefs conditional on the history $h_t$. In order to consistently estimate this bound, I appeal to the fact that at each instant a large number of traders will submit orders outside of the spread and thereby not affect the observable state of the book and not affect $h_t$. I use the cross-section of traders on both sides of the market to identify a bound on the distribution of conditional beliefs, $G(s|h_t)$. The estimator for the upper bound is
\begin{eqnarray}
G(s|h)^u=\frac{1}{|\mathcal{N}_t^a|}\sum_{i\in \mathcal{N}_t^a} \mathbf{1}\{p_{it}\leq s\} \mathbf{1}\{h_{it}=h\}
\end{eqnarray}
where $\mathcal{N}_t^a$ is the set of active traders at history $h_t$. In Figure \ref{ub_hist_4} I show estimates conditional on three different histories at times 40, 20, and 10 days before the end of the market with 95\% confidence bounds.\footnote{These are estimated by a subsampling routine outlined in the appendix. For further details on the estimation of confidence intervals in partially identified models see \citet*{tamer}.} Figure \ref{ub_hist_4} is for the Ted Cruz asset, in the appendix I show distributions for Trump and Rubio. If traders are approaching a consensus we would expect these distributions to be shifting towards a specific direction. 
%%%%%%%%%%%%%%%%%%
%%%%%%%%%%%%%%%%%%
These distributions look similar. To formally test this, I run a K-sample Kolmogorov-Smirnov test, which is another indication that orders are not systematically moving in a specific direction as resolution approaches. If consensus is not revealed publicly in the market, the a next question is whether an econometrician with access to more information than traders could act as an aggregator. The market has succeeded in attracting a large number of traders whose actions might reveal their private information in other dimensions. The next section studies whether this can be accomplished.
%%%%%%%%%%%%%%%%%%
\subsection{Leveraging Large Data on Orders}
I now consider whether an econometrician can act as an information aggregator. The data consist of a large number of realizations on order decisions, which potentially reveal the private signals of informed traders. An econometrician could conceivably identify the distribution of beliefs and then estimate the consensus belief of the market. With this in mind, I add further structure to facilitate the identification of private signals. The previous results indicate that private beliefs are not necessarily updated as time passes. This motivates my next assumption.  
%%%%%%%%%%%%%%%%%%
\begin{assumption}
Information contained in the book is not used to update private beliefs on the true probability of an outcome. Specifically:
\begin{eqnarray}
q_{it}(h_t)=q_i,\forall h_t, t
\end{eqnarray}
\end{assumption}
%%%%%%%%%%%%%%%%%%
This assumption is in line with the empirical evidence; the market has a large number of anonymous players, and the platform provides only limited information on the overall shape of the limit order book. Specifically, noise traders' activities are indistinguishable from those of  informed traders and conditioning beliefs on public information could lead to biased belief updates. The fact that public information is ``contaminated" by noise traders might support the existence of separating strategies of informed traders, which might partially reveal private information. This fact motivates imposing more structure on the order decision problem of an informed trader. A limit order is chosen if the expected payoff dominates the expected payoff from a market order. When a limit order of size $y_i$ is observed at price $p_t\in[0,m_t)$ where $m_t$ is the best ask at time $t$, it must be the case that:
\begin{eqnarray}
\phi(p_t,y_i)(q_{i}-p_t)\geq (q_{i}-m_t)
\end{eqnarray}
which leads to upper bound:\footnote{If there is not enough supply to execute only one market order at one price and a limit order is submitted then the following holds:
\begin{eqnarray}
\phi(p_t,y_i)(q_{i}-p_t)x_t\geq \sum_{k\in\sigma_{st}}(q_{i}-m_{kt})x_{kt}\mathbf{1}\left\{\sum_{l=1}^k x_{lt}\leq x_t\right\}
\end{eqnarray}
In other words, the expected payoff from the limit order is greater than ``walking-up" the book and submitting market orders until the demand $x_t$ is filled. This changes the upper bound on $q_i$:
\begin{eqnarray}
q_{i}\leq \frac{ \sum_{k\in\sigma_{1t}}m_{kt}x_{kt}\mathbf{1}\left\{\sum_{l=1}^k x_{lt}\leq x_t\right\}-p_tx_t}{1-\phi(p_t,y_i)}
\end{eqnarray}}
\begin{eqnarray}\label{eq:upper_bound}
 q_{i}\leq\frac{m_t-\phi_t(p_t,y_i)p_t}{1-\phi_t(p_t,y_i)}
\end{eqnarray}
Without any further assumptions, the upper bound on $q_{i}$ is not identified from the data. In particular, it is not possible to directly identify $\phi_t(p_t)$ without further discussion of equilibrium behavior. The next assumption allows for the identification of execution probabilities. 
%%%%%%%%%%%%%%%%%%%%%%%%%%%%%%%%%%%%%%%%%%%%%%%%%%%%%%%%%%%%%%%%%%%%%
\begin{assumption}
Traders ignore the history of the limit order book and use only the current best bid and ask to form beliefs on execution probabilities. That is,
\begin{eqnarray}\label{markov}\!\!\!\!\!\!\!\!\!\!\!\!\!\!\!\!\!\!\!\!\!\!\!\!\!\!\!\!\!\!\!\!\!\!\!\!\!\!\!\!\!\!\!\!\!\!\!\!\!\!\!\!\!\!\!\!\!\!\!\!\!\!\!\!\!\!\!\!\!\!\!\!
\Pr(\textrm{\emph{Order Executes at Price}}\  p_t\ \textrm{\emph{and quantity} } y_i|h_t)=
\end{eqnarray}
\begin{eqnarray}\qquad\qquad\qquad
\Pr(\textrm{\emph{Order Executes at Price}}\  p_t\ \textrm{\emph{and quantity} } x_t|\sigma_{0t},\sigma_{1t},\sigma_{0t}^x,\sigma_{1t}^x)\equiv\phi_t(p_t,x_t)
\nonumber\end{eqnarray}
\end{assumption}
In other words, previously executed trades provide no information on the future arrival of limit or market orders. Because traders are not updating beliefs based on the spread and the history of the spread, conditioning on this information will not provide information on the future spread. Execution thus depends on the arrival of new traders. The current spread provides partial information on how many orders ahead of the current submission must execute before trader $i$'s order is the first in line. We will denote this sub-vector $(\sigma_{0t},\sigma_{1t},\sigma_{0t}^x,\sigma_{1t}^x)$ by $\hat{h}_t$. This assumption is akin to assuming that probabilities are stationary Markovian with the current observable part of the book treated as state variables. When estimating these bounds, I exclude the last ten days of trade. I use all data on executions, including limit orders submitted by noise traders to estimate execution probabilities.\footnote{The fact that an order originated from a noise trader should not alter inferences on execution probabilities.} However, the bounds are estimated only for informed traders.

The lower and upper bounds on the belief distributions are shown including 95\% confidence bounds. In the Cruz and Trump market the belief bounds are very wide. The Cruz lower bound suggests that most beliefs are above 0.5, and the upper bound suggests that beliefs are split above and below 0.5. The 95\% confidence intervals also highlight how noisy these bounds are and that we cannot reject the possibility that the majority of traders have beliefs below 0.5. The converse can be seen in the market for Trump. Most beliefs are below 0.5; however, there is a reasonable amount of beliefs above 0.5 as well. The confidence intervals again show that it is possible that beliefs could be mostly above 0.5. In Table \ref{mean_beliefs} I show the estimated bounds on the mean beliefs implied by these distributions. In the case of Trump and Cruz the intervals include beliefs above and below 0.5. For Rubio beliefs are firmly below 0.5. These results suggest that the market did not provide any more information than what was public at the time. 
%%%%%%%%%%%%%%%%%%
\input{tables/mean_beliefs}
%%%%%%%%%%%%%%%%%%

%%%%%%%%%%%%%%%%%%%%%%%%%%%%%%%%%%%%%%%%%%%%%%%%%%%%%%%%%%%%%
\begin{figure}[h]
\centering\caption{Bounds on Private Information Distribution Ted Cruz Conditional on History, Iowa Caucus\label{ub_cdf}}
\includegraphics[width=.75\textwidth]{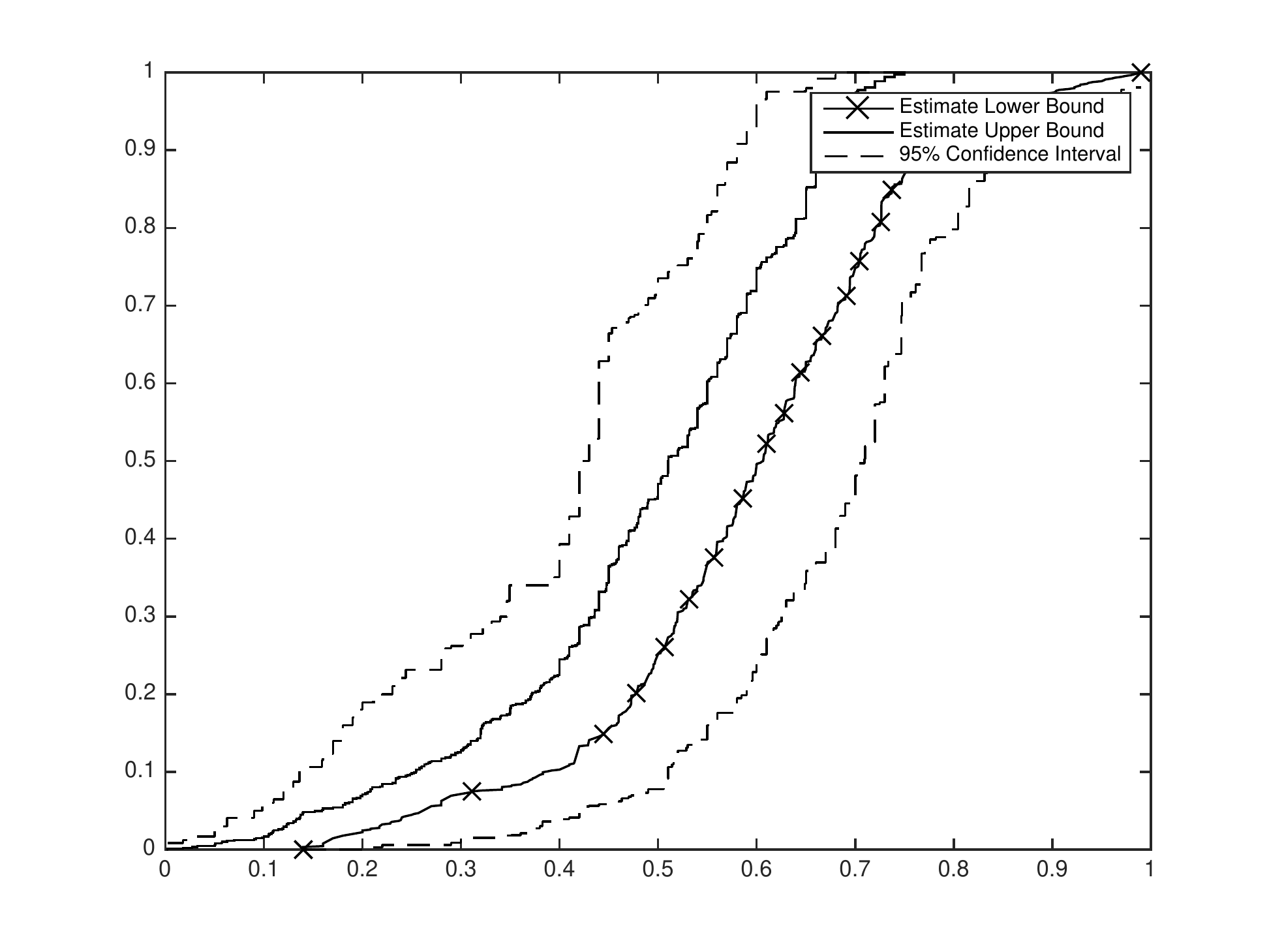}
\end{figure}
\begin{figure}[h]
\centering\caption{Bounds on Distribution of Conditional Beliefs Trump, Iowa Caucus}
\includegraphics[width=.75\textwidth]{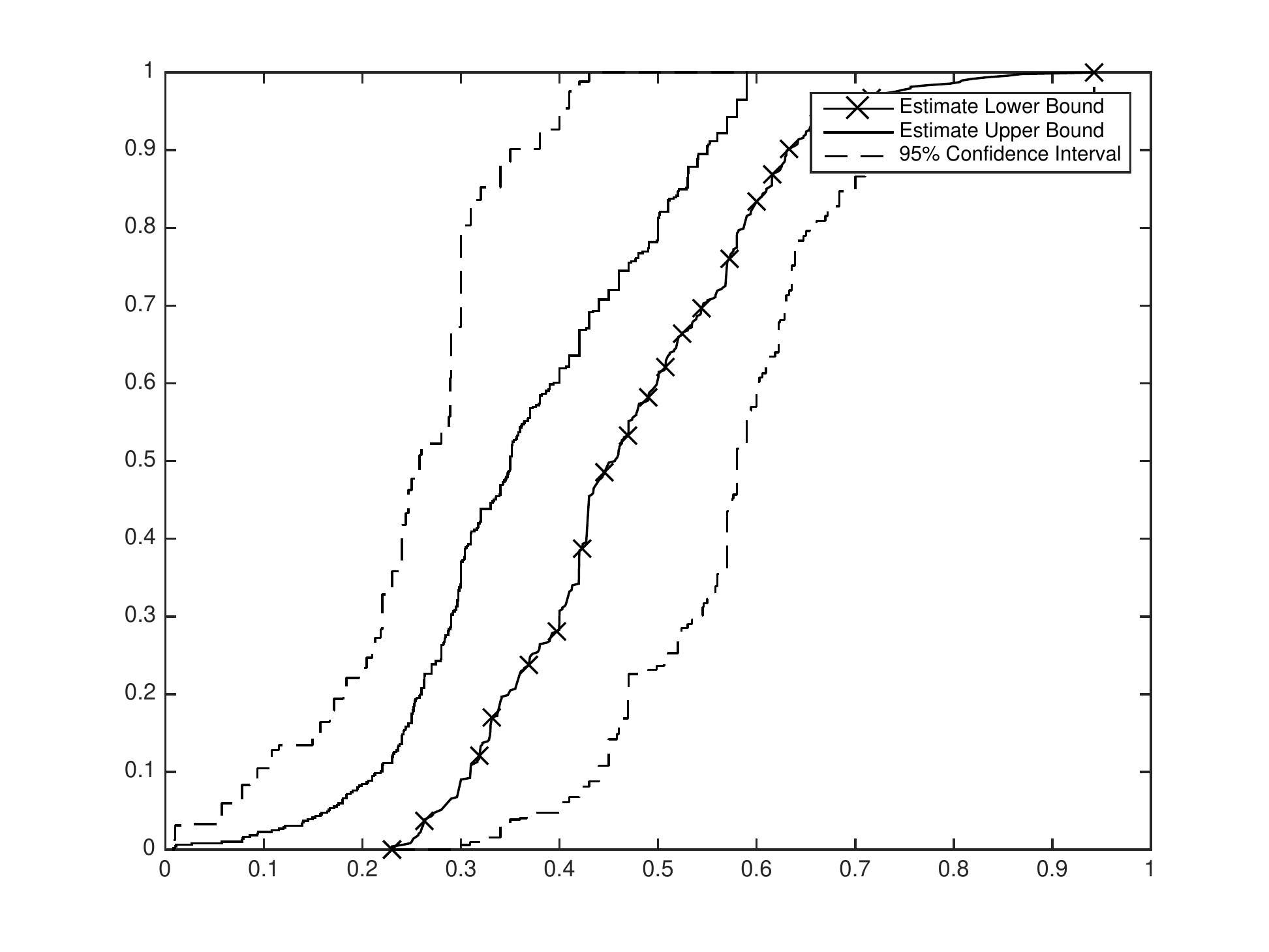}
\end{figure}

\begin{figure}[h]
\centering\caption{Bounds on Private Information Distribution Rubio Conditional on History, Iowa Caucus\label{ub_cdf}}
\includegraphics[width=.75\textwidth]{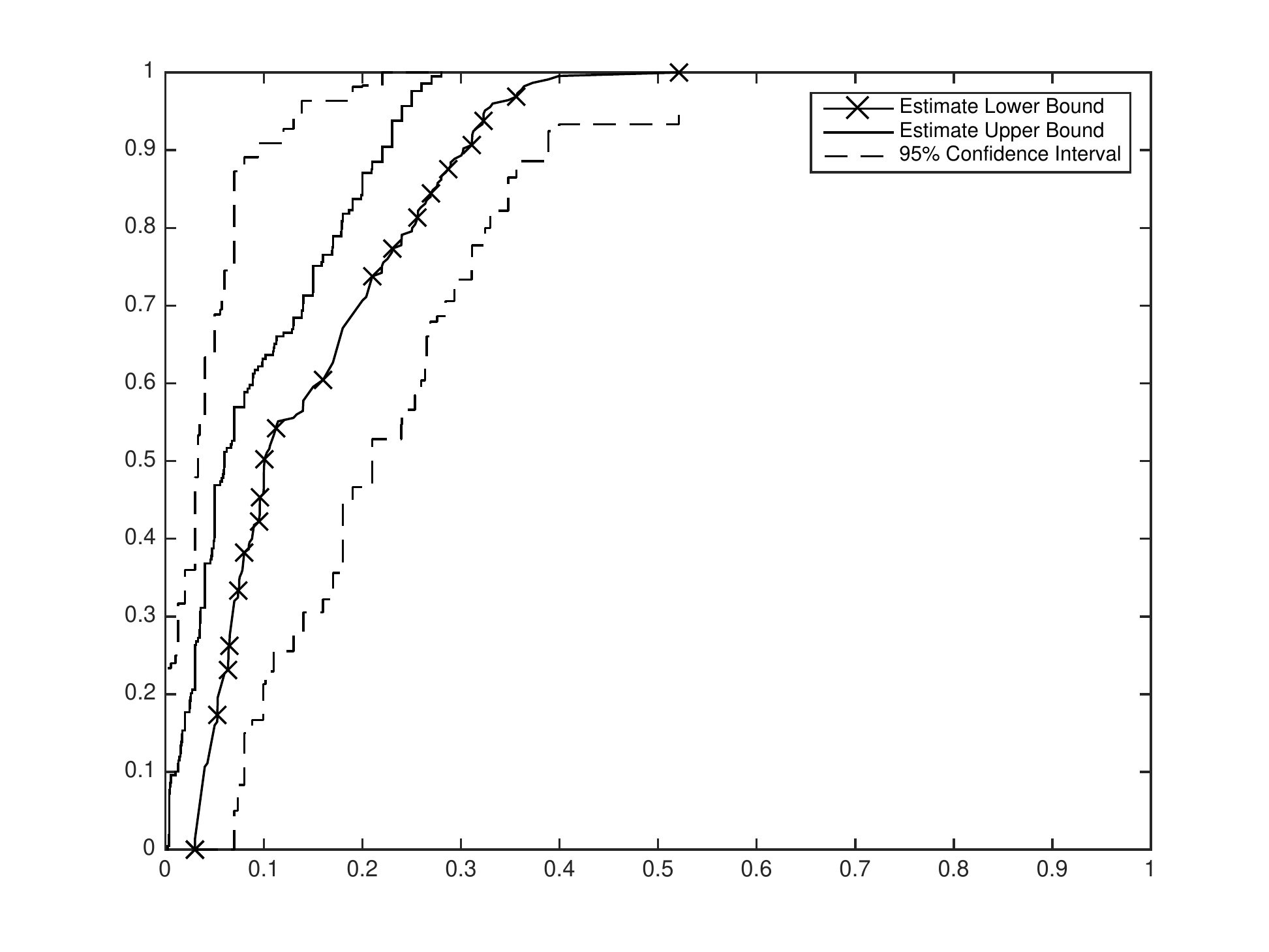}
%\centering\caption{Bounds on Distribution of Conditional Beliefs Trump, Iowa Caucus}
%\includegraphics[width=.75\textwidth]{img/yes_belief_TRUMPIACAUCUS16GOP_days_before_30}
\end{figure}

\clearpage

%%%%%%%%%%%%%%%%%%%%%%%%%%%%%%%%%%%%%%%%%%%%%%%%%%%%%%%%%%%%%
\section{Conclusion}
In this paper I empirically investigate information aggregation in a market organized around a limit order book. My goal is to determine whether the market converges on a consensus belief. The data suggest that beliefs are not updated and, therefore, disagreement amongst traders persists until the resolution of uncertainty. The evidence suggests that the presence of non-rational noise traders might be a key impediment to achieving a consensus, because their order submissions generate trading opportunities for informed traders to hide private information. 

I then consider whether an econometrician with access to the complete order book can leverage the large number of actions to aggregate signals, as opposed to ``outsourcing" this to the market. The results provide little support that this approach can generate informative estimates of mean beliefs. In other words, mean beliefs cannot be guaranteed to be different from 0.5. 

With these results, how do we salvage the promise of prediction markets? Developing mechanisms to exclude noise traders from public electronic markets is a potential conclusion of my study. Designing these systems to avoid excluding informed agents is hard to achieve.   Even if it were possible, the limit order market might not be the best mechanism. An informed trader would not want to reveal private information, i.e. play a separating equilibrium, which would allow later entrants to use previous orders to profitably update their beliefs. It is likely that little trade would be observed and a spike of activity would occur close to the resolution of uncertainty. The problem with this outcome is that prediction markets promise to provide better information before the resolution of uncertainty. 

This paper lends support to the assertion in \citet{shum} that a betting market might have better information aggregating properties than trading platforms. Their betting platform prevents the private information of one player to be used by another, which incentivizes participants to behave in a ``truth-telling" manner. The results in that paper are more promising than the findings of the large open platform I have studied.

\vspace{.1in}
\noindent TEPPER SCHOOL OF BUSINESS, CARNEGIE MELLON UNIVERSITY
%%%%%%%%%%%%%%%%%%%%%%%%%%%%
\newpage
\appendix
\begin{appendixnumbering}
\section{Appendix}
\subsection{Trade in Linked Securities}
Consider the following scenario: three political candidates $(A,B,C)$ are vying for public office. Only one can win. A trader believes that $A$ is going to win. The ``yes" asset for $A$ is trading at \$0.5 and the trader purchases six shares. The traders can also by ``no" assets for $B$ and $C$. $B$'s ``no" is trading at \$0.75 and $C$'s ``no" is trading at \$0.80. Payments are equal to:
\begin{table}[h]\centering\caption*{Multiple Asset Example}
\begin{tabular}{|c|c|c|c|}\hline
Asset&Price&Qty&Payment\\\hline\hline
A&0.50&6&3\\
B&0.75&4&3\\
C&0.80&5&4\\\hline
Total&&&10\\\hline
\end{tabular}
\end{table}
Payments in this market only need to equal the maximum exposure of the trader. Therefore, the payment necessary to cover these trades is not \$10 but something less. This is determined by considering the worst possible outcome for this position. The worst outcome involves getting two predictions wrong. The possible outcomes are:
\begin{enumerate}
\item A loses, B wins, C loses, total loss is $-3-3-4+5=1$
\item A loses, B loses, C wins, total loss is $-3-3+4-4=-6$
\end{enumerate}
and therefore payment will only equal $-6$.

%%%%%%%%%%%%%%%%%%%%%%%%%%%%%%%%%%%%%%%%%%%%%%%%%%%%%%%%%%%%%%%%%%%%%%%%%%%%%%%%%%%%%%%%%%%%%%%%%%%%%%%%%

\section{Further Descriptive Analyses for the Iowa Caucus}
%%%%%%%%%%%%%%%%%%%%%%%%%%%%%%%%%

\subsection{Profits and their Relationship to Trading Behavior}
I now seek to determine the link between trader behavior and their terminal profits. I will break down the set of traders into the two groups, the top 30 traders by profits and the bottom 30. I first consider if the order times for traders are linked to their final profitability. This is shown in Figure \ref{order_time}. The major difference between the two distributions is the greater mass between 0.8 and 0.9 for the more profitable traders, suggesting that the successful traders entered earlier, however, this is a very small difference. 
%%%%%%%%%%%%%%%%%%%%%%%%%%%%%%%%%
%\begin{figure}[h]
%\centering\caption{Empirical Distribution of Order Time\label{order_time}}
%\includegraphics[width=.75\textwidth]{img/order_time_top}
%\end{figure}
%%%%%%%%%%%%%%%%%%%%%%%%%%%%%%%%%
I next look at execution times for the top and bottom traders. Successful traders are trading slightly faster than the bottom traders. The higher y-intercept for the top traders indicates that successful traders are more likely to submit a market order that executes immediately. 
%%%%%%%%%%%%%%%%%%%%%%%%%%%%%%%%%
%\begin{figure}[h]
%\centering\caption{Empirical Distribution of Execution Time\label{exe_time_top}}
%\includegraphics[width=.75\textwidth]{img/exe_time_top}
%\end{figure}
%

%%%%%%%%%%%%%%%%%%%%%%%%%%%%%%%%%

\subsection{Portfolio Shifts for Top 30 Traders}
I now consider the same analysis for the top traders as measured by their profits from accurate predictions. I focus on the Iowa caucus market, since there were many securities and many ways of being correct, i.e. by taking a position against a candidate's success. I now show the portfolio shift matrix for the top 30 traders by prediction profits. The picture is very similar to the complete set of traders. In particular, successful traders started with securities in favor of Cruz and when possible purchased more. 

\subsection{Upper Bounds on Conditional Signal Distribution}
In this section, I show the upper bound for the belief distribution. The results are similar to the Cruz market, where distributions shift little and the hypothesis that these distributions are all from the same parent distribution cannot be rejected at 99\%. 
%\begin{figure}[h]\centering\caption{Upper Bound on Distribution of Conditional Beliefs Trump, Iowa Caucus}
%\includegraphics[width=.75\textwidth]{img/ub_hist_13}
%\end{figure}
%\begin{figure}[h]
%\centering\caption{Upper Bound on Distribution of Conditional Beliefs Trump, Iowa Caucus}
%\includegraphics[width=.75\textwidth]{img/ub_hist_13}
%\end{figure}
%
%\begin{figure}[h]
%\centering\caption{Upper Bound on Distribution of Conditional Beliefs Rubio, Iowa Caucus}
%\includegraphics[width=.75\textwidth]{img/ub_hist_11}
%\end{figure}
%
%\clearpage
%%%%%%%%%%%%%%%%%%%%%%%%%%%%%%%%%%%%%%%%%%%%%%%%%%%%%%%
%%%%%%%%%%%%%%%%%%%%%%%%%%%%%%%%%%%%%%%%%%%%%%%%%%%%%%%
%\subsection{Informal Evidence of Stationarity of Prices}
%
%\begin{figure}[h]\centering
%\caption{Yes Price Path Across Trading Days Excluding Last 30 Trading Days, Marriage Equality}
%\includegraphics[width=.75\textwidth]{img/time_yes_price}
%\end{figure}
%
%\begin{figure}[h]\centering\caption{Empirical Distribution of Across Trading Days Excluding Last 30 Trading Days, Marriage Equality}%\label{avg_price_sup}}
%\includegraphics[width=.75\textwidth]{img/cdf_yes_price}
%\end{figure}
%\clearpage
%

%%%%%%%%%%%%%%%%%%%%%%%%%%%%%%%%%%%%%%%%%%%%%%%%%%%%%%%
\newpage
\section{Marriage Equality Market}
In the marriage equality market, there were total of 1211 active traders.  On average across days, the average number of active traders is 25.5. In this market the asset traded on average at a price \$0.75. The average price path is shown in Figure \ref{avg_price_sup}. The event in question did occur and therefore this could provide the first piece of evidence that execution prices could be used as sufficient statistics for the true probability an event is going to occur. The problem with this interpretation is due to the market clearing mechanism. As mentioned previously, the market requires disagreement amongst its traders to enable exchange. Given that positive trade is observed this has to be the case. However, there is no market maker, who ensures that demand for all assets is satisfied as much as possible. In Figure \ref{volume_sup} the daily volume of trade is shown. Volumes move around an average of 500 securities a day with a large spike as the event approaches. If these markets are to provide early information aggregation this volume spike is undesirable. 
%%%%%%%%%%%%%%%%%%%%%%%%%%%%%%%%%%%%
%\input{tables/summary_sup}
%%%%%%%%%%%%%%%%%%%%%%%%%%%%%%%%%%%%
%
%\begin{figure}[h]
%\centering\caption{Execution Prices Supreme Court\label{avg_price_sup}}
%\includegraphics[width=.75\textwidth]{img/p_sup}
%\end{figure}
%\begin{figure}[h]
%\centering\caption{Daily Volume Supreme Court\label{volume_sup}}
%\includegraphics[width=.75\textwidth]{img/vol_sup}
%\end{figure}
%
%%%%%%%%%%%%%%%%%%%%%%%%%%%%%%%%%

\subsection{Portfolio Shifts}
In the case of the Supreme Court ruling there are three possible positions, in favor, against or a zero portfolio position. Darker shades indicate a higher count. In Figure \ref{port_shift_sup} most of the entries are on the diagonals, suggesting that traders start with a Yes or No asset and if they do shift change their holdings they are mostly staying in the same asset class. In Figure \ref{port_shift_sup} we can see that most activity was in the ``No" asset.
%
%\begin{figure}[h]\centering\caption{Portfolio Shifts, Marriage Equality\label{port_shift_sup}}
%\includegraphics[width=.75\textwidth]{img/pos_shift_hi_time_sup_7361416e+05.pdf}
%\end{figure}
%
%\begin{figure}[h]\centering\caption{Portfolio Shifts, Marriage Equality\label{port_shift_sup}}
%\includegraphics[width=.75\textwidth]{img/pos_shift_hi_time_sup_7361305e+05.pdf}
%\end{figure}
%\clearpage 
%\subsection{ECDFs of Profits by Entry Time}
%\begin{figure}[h]
%\centering\caption{Empirical Distribution of Profits by Entry Time, Marriage Equality\label{pi_time_sup}}
%\includegraphics[width=.75\textwidth]{img/profit_time_sup}
%\end{figure}
%
\clearpage
%%%%%%%%%%%%%%%%%%%%%%%%%%%%%%%%%%%%%%%%%%%%%%%%%%%%%%%

\section{Details on Estimation}
\subsection{Execution Probabilities}
In order to estimate the execution probabilities I use a Nadaraya-Watson estimator. In particular:
\begin{eqnarray}
\widehat{\phi}(p)=\frac{\sum_{t=1}^T \sum_{i=1}^N\mathbf{1}\{\textrm{order executes}\}K_h(x_{it}-x)k_{h_p}(p_{it}-p)}{\sum_{t=1}^T \sum_{i=1}^NK_h(x_{it}-x)k_{h_p}(p_{it}-p)}
\end{eqnarray}
where $x=\hat{h}$. We use the product of normal densities as the multivariate kernel, specifically
\begin{eqnarray}
K_h(x_{it}-x)=\prod_{m=1}^M \phi\left(\frac{x_{itm}-x_m}{h_m}\right)
\end{eqnarray}
where $M$ is the dimensionality of $\hat{h}$ and $h_m$ is a bandwidth chosen using a multivariate Silverman's Rule. 

\subsection{Subsampling}

To estimate confidence intervals, I use a subsampling procedure where I re-sample trading days and repeatedly estimate the parameters of interest. This procedure is described in \citet*{politis} and is shown to be more robust than the standard bootstrap. I describe the general approach below. We first describe how we approximate the sampling distribution of the estimator. A standard subsampling algorithm uses subsamples of size $b<<A$ to re-estimate the parameters of interest, where $A$ is the number of auctions. Denote our parameter estimate by $\widehat{\theta}_A$ and the parameter estimate from subsample $a$ of size $b$ by $\widehat{\theta}_{A,b,a}$. If we create $q$ subsamples and re-estimate the parameter on these subsets we can then use the following as an approximation for the sampling distribution: 
\begin{eqnarray}
L_{A,b,||.||}(x|\tau_b)\equiv q^{-1}\sum_{a=1}^q\mathbf{1}\left\{\tau_b||%
\widehat{\theta}_{A,b,a}-\widehat{\theta}_A||\leq x\right\}
\end{eqnarray}
where $\tau_b$ is the convergence rate of the estimator and $||.||$ is the norm of the normed linear space $\Theta$. Theorem 2.5.2 in \citet*{politis} establishes the asymptotic validity of this approach. $q$ represents the number of all possible combinations of the data, i.e. $\left(\begin{array}{c}A \\ b\end{array}\right)$. This can easily be an incredibly large number, \citet{politis} therefore also prove that a stochastic approximation leads to consistent estimates by randomly selecting observations with or without replacement from the set of trading periods $\{1,\dots,A\}$ which substantially reduces the computational burden. I run 500 subsamples to estimate confidence intervals. 

\vspace{.1in}
\noindent \emph{Estimating Convergence Rates}
If we have, two different block sizes, we can subtract two
versions of the above for the two different block sizes. In addition, if we
assume that $\tau_n=n^\beta$, we can state: 
\begin{eqnarray}
\log\left(\frac{b_1}{b_2}\right)^{-1}\left( \log L^{-1}_{A,b_2}(s|1)- \log
L^{-1}_{A,b_1}(s|1)\right)=\beta+o_P\left(\log\left(\frac{b_1}{b_2}%
\right)^{-1}\right)
\end{eqnarray}
It is possible to estimate $\beta$ using more than two block sizes. Let 
\begin{eqnarray}
y_{k,j}\equiv
\log\left(L^{-1}_{A,b_k,||.||}(s_j|1)\right)=a_k-\beta\log(b_k)+u_{k,j}
\end{eqnarray}
where $a_j\equiv\log\left(J^{-1}_{||.||}(s_j,P) \right)$ and $u_{k,j}=o_P(1)$%
. In particular, if we have $K$ different block sizes we can estimate $\beta$
as follows: 
\begin{eqnarray}
\beta=\frac{\sum_{k=1}^K(y_k-\overline{y})(\log(b_k)-\overline{\log})}{%
\sum_{k=1}^K(\log(b_k)-\overline{\log})^2}
\end{eqnarray}
where 
\begin{eqnarray}
y_k=J^{-1}\sum_{j=1}^J y_{k,j}
\end{eqnarray}
\begin{eqnarray}
\overline{y}=(KJ)^{-1}\sum_{k=1}^K\sum_{j=1}^J y_{k,j}
\end{eqnarray}
\begin{eqnarray}
\overline{\log}=K^{-1}\sum_{k=1}^K\log(b_k)
\end{eqnarray}
The asymptotic validity of this approach is established in Theorem 8.2.3. in \citet{politis}.  

\end{appendixnumbering}
\clearpage

%%%%%%%%%%%%%%%%%%%%%%%%%%%%%%%%%%%%%%%%%%%%%%%%%%%%%%
\bibliographystyle{plainnat}

\bibliography{references}
%%%%%%%%%%%%%%%%%%%%%%%%%%%%%%%%%%%%%%%%%%%%%%%%%%%%%%
%\newpage
%\section{Tables}
%\input{tables}
%%%%%%%%%%%%%%%%%%%%%%%%%%%%%%%%%%%%%%%%%%%%%%%%%%%%%%%
%
%\newpage
%\section{Figures}
%\input{figs}
\end{document}

%% file: tables/summary.tex
\begin{table}[h]\caption{Summary Statistics\label{sum_stat}}\centering
{\begin{tabular}{lcccc}
\hline\hline
&\textbf{Mean}&\textbf{Std}&\textbf{Min}&\textbf{Max}\\\hline
{CRUZ Average Daily Price}&0.45&0.12&0.31&0.67\\
{CRUZ Daily Volume}&1006.57&3254.08&0.00&29928.00\\
{CRUZ Daily Number of Trades}&21.74&59.52&0.00&548.00\\
{CRUZ Daily Number of Traders}&11.11&86.98&0.00&2400.00\\
{CRUZ Average Terminal Holdings Yes}&1.28&31.33&0.00&1330.00\\\hline
{RUBIO Average Daily Price}&0.07&0.01&0.05&0.15\\
{RUBIO Daily Volume}&858.92&3886.91&0.00&51334.00\\
{RUBIO Daily Number of Trades}&11.90&44.10&0.00&575.00\\
{RUBIO Daily Number of Traders}&7.95&123.88&0.00&3840.00\\
{RUBIO Average Terminal Holdings Yes}&3.42&47.12&0.00&928.00\\\hline
{TRUMP Average Daily Price}&0.62&0.10&0.30&0.73\\
{TRUMP Daily Volume}&1195.88&4059.40&0.00&51834.00\\
{TRUMP Daily Number of Trades}&24.45&76.16&0.00&945.00\\
{TRUMP Daily Number of Traders}&31.27&168.92&0.00&3199.00\\
{TRUMP Average Terminal Holdings Yes}&27.72&154.92&0.00&2445.00\\\hline
{Total Number of Traders}&4452
\end{tabular}}
\end{table}

%% file: tables/profits.tex
\begin{table}[h]\caption{Summary Statistics on Profits\label{profits}}\centering
{\begin{tabular}{lcccc}
\hline\hline
&\textbf{Mean}&\textbf{Std}&\textbf{Min}&\textbf{Max}\\\hline
{Prediction Profits Iowa Caucus}&16.46&92.00&-486.25&1607.20\\
{Trading Profits Iowa Caucus}&-214.71&800.73&-9347.12&1469.49\\
{Trading Profits Iowa Caucus, DT}&-27.14&253.16&-2314.11&2899.79\\
\end{tabular}}
\end{table}

%% file: tables/mean_beliefs.tex
\begin{table}[h]\caption{Mean Beliefs Implied by Bounds\label{mean_beliefs}}\centering
{\begin{tabular}{lccccc}
\hline\hline
&\multicolumn{2}{c}{\textbf{Lower Bound 95\%}}&\multicolumn{2}{c}{\textbf{Upper Bound 95\%}}&\textbf{Average}\\[-.5ex]
&\multicolumn{2}{c}{\textbf{Confidence Interval}}&\multicolumn{2}{c}{\textbf{Confidence Interval}}&\textbf{Transaction Price}\\\hline
{CRUZ}&0.41&0.54&0.50&0.65&0.52\\
{RUBIO}&0.06&0.13&0.10&0.19&0.12\\
{TRUMP}&0.26&0.46&0.38&0.56&0.39\\
\end{tabular}}
\end{table}